\documentclass[structabstract]{aa}  
\usepackage{graphicx, natbib, latexsym,amssymb,amsmath}
\usepackage{txfonts}
\bibpunct{(}{)}{;}{a}{,}{,}
\usepackage{supertabular}
\usepackage{longtable,lscape}
\def\aap{A\&A}                
\def\apjl{ApJ}                
\def\apjs{ApJS}               
\def\amm{$\mathrm{NH_3}$}
\def\hcop{$\mathrm{HCO^+}$}
\def\hh13cop{$\mathrm{H^{13}CO^+}$}
\def\n2hp{$\mathrm{N_2H^+}$}
\def\c18o{$\mathrm{C^{18}O}$}
\def\mecn{$\mathrm{CH_3CN}$}
\def\h2co{$\mathrm{H_2CO}$}
\def\arcsec{$''$}
\def\arcmin{$'$}
\def\degr{$^\circ$}

\begin{document}

   \title{Initial phases of massive star formation in high infrared extinction
   clouds}

   \subtitle{II. Infall and onset of star formation \thanks{This publication is based on data acquired with the IRAM 30m Telescope and the Atacama Pathfinder Experiment (APEX). IRAM is supported by INSU/CNRS (France), MPG (Germany), and IGN (Spain). APEX is a collaboration between the Max-Planck-Institut f\"ur Radioastronomie, the European Southern Observatory, and the Onsala Space Observatory.}}
   \author{K.~L.~J. Rygl\inst{1,2},
          F. Wyrowski\inst{2},
          F. Schuller\inst{3,2} \and K.~M. Menten\inst{2}
          }
   \institute{ Istituto di Astrofisica e Planetologia Spaziali (INAF-IAPS), 
   Via del Fosso del Cavaliere 100, 00133 Roma, Italy\\
              \email {kazi.rygl@inaf.it}
   \and Max-Planck-Institut f\"ur Radioastronomie (MPIfR),
              Auf dem H\"ugel 69, 53121 Bonn, Germany\\
              \email{[wyrowski;kmenten]@mpifr-bonn.mpg.de}
   \and  European Southern Observatory (ESO), Alonso de Cordova 3107, Casilla 19001, Santiago 19, Chile\\
              \email{fschulle@apex-telescope.org}           
              }
   \date{}

 
  \abstract
   {The onset of massive star formation is not well understood because of observational and theoretical difficulties. To find the dense and cold clumps where massive star formation can take place, we compiled a sample of high infrared extinction clouds. We observed the clumps in these high extinction clouds in the 1.2\,mm continuum emission and ammonia with the goals of deriving the masses, densities, temperatures, and kinematic distances.}
   {We try to understand the star-formation stages of the high extinction clumps by studying their infall and outflow properties, the presence of a young stellar object (YSO), and the level of the CO depletion. Are the physical parameters, density, mass, temperature, and column density correlated with the star-forming properties? Does the cloud morphology, quantified through the column density contrast between the clump and the clouds, have an impact on the evolution of star formation occurring inside it?}
   {Star-formation properties, such as infall, outflow, CO depletion, and the presence of YSOs, were derived from a molecular line survey performed with the IRAM 30m and the APEX 12m telescopes.}
    {We find that the \hcop(1--0) transition is the most sensitive for detecting infalling motions. SiO, an outflow tracer, was mostly detected toward sources with infall, indicating that infall is accompanied by collimated outflows. We calculated infall velocities from the line profiles and found them to be of the order of 0.3--7\,$\mathrm{km~s^{-1}}$. The presence of YSOs within a clump depends mostly on the clump column density;  no indication of YSOs were found below $4\times10^{22}\,\mathrm{cm^{-2}}$. 
    }
   {Star formation is on the verge of beginning in clouds that have a low column density contrast; the infall is not yet present in the majority of the clumps. The first signs of ongoing star formation are broadly observed in clouds where the column density contrast between the clump and the cloud is higher than two; most clumps show infall and outflow. Finally, we find the most evolved clumps in clouds that have a column density contrast higher than three; in many clumps, the infall has already halted, and toward most clumps we found indications of YSOs. Hence, the cloud morphology, based on the column density contrast between the cloud and the clumps, seems to have a direct connection with the evolutionary stage of the objects forming inside. }

   \keywords{ Stars: formation -- ISM: clouds -- ISM: jets and outflows -- ISM: molecules  -- Submillimeter: ISM
                 --               
               }
\authorrunning{K.~L.~J. Rygl et al.}
\titlerunning{Infall and onset of star formation}

   \maketitle

\section{Introduction}
\label{sec:one}
Massive star formation is thought to occur in clumps that are deeply embedded in the dense gas and dust of the natal molecular cloud. In these cold and dense environments, observations are often hindered by absorption or high optical depth of the emission and the confusion of many objects due to a limited angular resolution. Molecular transitions in the submm wavelength range are particularly useful probes to study these regions, since their critical densities are similar to the densities one expects for regions in which stars are forming. Even without resolving the star-forming clumps, one can find signposts of star formation, such as outflows and infall, from the kinematics derived from the shapes and radial velocities of molecular lines. In this paper, we try to use the molecular emission observed toward the sample of high extinction clouds, whose
continuum and ammonia observations were discussed in \citet{rygl:2010b}, hereafter Paper I,  to study the infall properties of the clumps and other indications for the onset of star formation.

Crucial for the understanding of massive star formation is observational evidence for the infall of matter. In low-mass star-forming regions (SFRs) such as the Bok globule B335, infall has been inferred from the observed line profiles (\citealt{zhou:1993}). There are but few
infall studies for candidate high-mass stars (\citealt{wu:2005,fuller:2005,beltran:2006a}); however, their number is
currently increasing rapidly. A recent study by \citet{sepulcre:2010}, using a sample of infrared dark and infrared bright clumps from the literature, found a good correlation between the clump dust mass and the outflow mass, irrespective of the clump being infrared dark or bright. 
These authors claim that the outflow properties during star formation remain unchanged within a large range of masses and infrared properties of the clumps. 

For our infall study of dense clumps, we used not only the shapes of the \hcop(1--0) line but also the shapes of the \hcop(4--3), CO(3--2), and CO(1--0) emission lines. The different molecular transitions probe different densities and therefore different parts of the clump. Lower-$J$ CO lines trace the kinematics of the lower density material of the clumps. At high densities, $n\geq10^{4}\mathrm{cm^{-3}}$, and cold temperatures, $\sim$15\,K, the CO molecule depletes from the gas phase (\citealt{kramer:1999}). The level of CO depletion therefore indicates if the clump is cold and dense, and hence young, or if it consists of already more processed material. Apart from the presence of broad wings in lines from CO and other molecules, outflows can additionally be traced by an enhanced abundance of SiO, which is  a good shock tracer (\citealt{schilke:1997}). We also observed  \h2co\ and \mecn , which trace the more evolved stage of the young stellar object (YSO). 

Star formation is also evident by the presence of heating sources: a YSO's radiation increases the temperature of the dust and gas surrounding it. This is not only visible in infrared emission of the dust but also in several molecules like \h2co\ and \mecn . These molecules are often observed toward hot molecular cores, because the elevated temperatures ($>100$\,K) in these regions evaporate the icy grain mantles, increasing the gas phase abundances of such species by orders of magnitude (\citealt{mangum:1993, olmi:1996,tak:2000}). Additionally, for \mecn\ one can estimate rotational temperatures of the hot phase from the emission in lines arising from different $K$ levels that cover a wide range in temperature, but are close together in frequency.
Using these tracers, we define starless clumps as clumps without YSOs and signs of infall/outflows and prestellar clumps as clumps without YSOs but with infall/outflows.

The higher density components of the clumps were studied with \n2hp\ and \hh13cop; \n2hp\ is known to be a reliable probe of cold gas with lower
depletion than most other species (\citealt{tafalla:2002}). The hyperfine structure (hfs) of this
molecule also allowed to estimate its excitation temperature and column density. The optically thin \hh13cop(1--0) line was used to determine the local standard of rest velocity ($v_\mathrm{LSR}$) of the clump. 

We tried to correlate star-formation behavior in the clump not only with surface density but also with the morphology and physical conditions of the cloud.
In Paper I, we selected a sample of high extinction clouds based on our extinction maps from the first Galactic quadrant and performed 1.2\,mm dust continuum observations of them.
We defined three categories of clouds based on their 1.2\,mm dust continuum contrast between the clump column density and the cloud column density. Clouds with a low clump to cloud column density contrast ($C_{N_\mathrm{H_2}}$) $C_{N_\mathrm{H_2}}$$<$2 were defined as {\em diffuse clouds} and thought to be the youngest clouds: they are the coldest and show few signs of star formation, such as masers and infrared emission. Clouds with higher contrast 2$<$$C_{N_\mathrm{H_2}}$$<$3 were defined as {\em peaked clouds} having at least one clump (or peak, hence `peaked' clouds) with a high contrast to the cloud emission; these clouds were thought to be the following stage. They show slightly higher temperatures and wider line widths (larger turbulence). The most evolved clouds were the {\em multiply peaked clouds}, defined by a still higher contrast $C_{N_\mathrm{H_2}}$$>$3 and generally containing several clumps (hence `multiply peaked' clouds). Most of the maser emission and infrared sources were found toward these clouds. Even though several clouds, mostly the peaked and multiply peaked ones, are elongated or filamentary, this phenomenological feature was not taken into account in defining the three cloud categories.

The high extinction cloud sample that was studied in Paper I contained 50 clumps, of which 12 were in diffuse clouds, 19 in peaked clouds, and 19 in multiply peaked clouds. Almost all clumps showed \amm(1,1) emission, indicating the presence of high-density gas (Paper I). In the current paper, we report the results of a molecular survey on the clumps in high extinction clouds. With these results, we try to pinpoint their evolutionary stage and test our proposed evolutionary sequence of the clouds.
We observed all the positions that contained \amm(1,1) emission with the IRAM 30m telescope in the \hh13cop(1--0), SiO(2--1), \hcop(1--0), \mecn(5--4) $K$=0, 1, 2, 3, and 4 levels, \n2hp(1-0), \c18o(2--1), and CO(2--1) molecular transitions. Next, a subsample of these sources was observed with the Atacama Pathfinder Experiment submillimeter telescope (APEX) in higher $J$ transitions: \n2hp(3--2), \h2co(4$_{03}$--$3_{04}$), \hcop(4--3), and CO(3--2). The APEX targets included mostly active and evolved sources, which are expected to emit strongly in these higher excitation transitions. In addition, a few diffuse and peaked clouds were added to allow comparison between the three categories of clouds. The line data were interpreted using the RADEX radiative transfer code (\citealt{tak:2007}) to arrive at models of the physical parameters of the clumps. 

The results of the line observations are compared with results
from similar surveys toward other samples, e.g.\ the study of line
emission toward Infrared Dark Clouds (IRDCs) and other massive star-forming regions by \citet{ragan:2006}, \citet{motte:2007}, \citet{pillai:2007}, \citet{pirogov:2007}, \citet{purcell:2009}, \citet{sepulcre:2010}, and \citet{vasyunina:2011}; the survey of methanol
maser sources by \citet{purcell:2006}; the high-mass protostellar
objects survey by \citet{fuller:2005} and \citet{beuther:2007b}; and similar work toward hot
molecular cores by \citet{araya:2005} and UCH{\sc ii}\,regions by \citet{churchwell:1992}. This comparison will allow assessment of the differences in evolutionary stages covered by these studies.

The observations and data reduction are described in Sect.~2. Sect.~3 gives
the results of the infall study, the derived temperature and density
estimates, the search for YSO indications, and the CO depletion
study. These results are interpreted in light of an evolutionary sequence of
the three classes of clouds in the discussion (Sect.~4) and summarized in Sect.~5.

\section{Sample of high extinction clumps and observations}
\subsection{The sample}
The high extinction clouds were selected from color-excess maps in the 3.6\,$\mu$m--4.5\,$\mu$m {\it Spitzer} IRAC bands (\citealt{fazio:2004}). The maps cover Galactic longitudes $10^\circ<l<60^\circ$ in the first quadrant, $295^\circ<l<350^\circ$ in the fourth quadrant, and $-1^\circ<b<1^\circ$ in Galatic latitude. The method to construct the extinction maps is described in Paper I. 
We selected these clouds based on a 3.6\,$\mu$m--4.5\,$\mu$m color excess (CE) above 0.25\,mag, which corresponds to a visual extinction 20.45\,mag by $A_V=81.8\times CE(3.6\,\mu\mathrm{m}-4.5\,\mu\mathrm{m})$ using the reddening law of \citet{indebetouw:2005} (for more details, see Paper I). 
To guarantee visibility with the Effelsberg 100m and the IRAM 30m telescopes, we focused on the clouds in the first Galactic quadrant. In Paper I we analyzed the dust continuum emission and the ammonia inversion transitions of the clumps in high extinction clouds to derive their physical properties: the clumps have masses between 10 up 400\,$M_\odot$, they are cold, $\sim$16\,K, and are found at distances between 1 and 7\,kpc, with the majority being at 3\,kpc. A summary of the properties of the diffuse, peaked, and multiply peaked high extinction clouds, based on Paper I, is presented in Table \ref{ta:sample}. Additionally, we calculated the surface density, following \citet{sepulcre:2010} Eq. 1: $\Sigma  = 4 M / \pi D^2$, where $M$ is the clump mass in grams and $D$ the diameter of the clump in cm.

\begin{table*}
\centering
\caption{Properties of clumps in high extinction clouds based on Paper I\label{ta:sample}}
\begin{tabular}{llccccccc}
\noalign{\smallskip}
\hline\hline
\noalign{\smallskip}
&& \multicolumn{2}{c}{diffuse clouds}& \multicolumn{2}{c}{peaked clouds}&\multicolumn{2}{c}{multiply peaked clouds}& all clouds\\
&& mean& \multicolumn{1}{c}{range}& mean& \multicolumn{1}{c}{range}& mean& \multicolumn{1}{c}{range}& mean\\
\noalign{\smallskip}
\hline
\noalign{\smallskip}
cloud mass   &$M_\odot$ &495 & 17 -- 3039 & 1420 & 70 -- 5500& 1900 & 500 -- 6500 & 910\\
cloud diameter     &pc& 0.8 & 0.2 -- 2.3& 1.7 & 0.7 -- 3.7 & 1.9 & 1.2 -- 3.8& 1.2 \\
\noalign{\smallskip}
clump mass  &$M_\odot$&105 & 12 -- 283 & 130& 28 -- 738& 185 & 21 -- 431 & 150 \\
clump $N_\mathrm{H_2}$ &$10^{22}\,\mathrm{cm^{-2}}$& 5.4 & 2.9 -- 9.5 & 5.5 &3.3 -- 7.7& 8.5& 2.5 -- 26.3 &6.8\\
clump diameter   &pc    &0.3 & 0.11 -- 0.56 & 0.33 & 0.14 -- 0.76 &0.36 & 0.15 -- 0.47& 0.3\\
clump $\Delta v$ &$\mathrm{km~s^{-1}}$& 1.2 & 0.7 -- 2.3&1.4 & 0.9 -- 2.5 &1.6 & 0.8 -- 2.8 & 1.5\\
clump d &kpc&2.8 &1.1 -- 5.7& 3.4 & 1.9 -- 7.2& 3.6 & 2.1-- 4.7 & 3 \\
clump temperature & K &13.5 & 9.3 -- 16.7 & 15.7 & 11.9 -- 18.6 & 17.5 & 12.4 -- 24.7 & 16\\
clump $\Sigma$ &$\mathrm{g~cm^{-2}}$&0.3 &0.2 -- 0.5 & 0.3 & 0.2 -- 0.5& 0.5 & 0.2 -- 1.5&0.4\\
\noalign{\smallskip}
\hline
\hline
\end{tabular}
\tablefoot{Rows are (from top to bottom) cloud mass, cloud diameter, clump mass, clump column density, clump diameter, clump line width, clump kinematic distance, clump rotational \amm\ temperature, and clump surface density.
}
\end{table*}

\subsection{IRAM 30m observations}

The spectral line survey of the clumps in high extinction clouds was performed
with the IRAM 30m telescope in 2007, June. 
We exploited the possibility of the ABCD receivers to observe simultaneously at two frequencies,
one at $\sim$100\,GHz and the other at $\sim$230\,GHz. With two different
receiver setups, we observed a total of seven molecular lines, listed in Table \ref{ta:freq}. The SiO(2--1) and \hh13cop(1--0) transitions were observed simultaneously in the same backend, since they are only separated by 100\,MHz. The higher frequency  CO(2--1) and \c18o(2--1) transitions were observed in both setups, which increased the signal-to-noise ratio by $\sqrt{2}$. 
For the $\sim$100\,GHz lines, we used several VESPA backend settings: one with a (relatively) narrow bandwidth of 40\,MHz and a channel spacing of 0.02\,MHz, corresponding to velocity resolution of 0.08\,km~s$^{-1}$, and two wider bandwidths of 120 and 160\,MHz using a channel spacing of 0.04\,MHz, corresponding to a velocity resolution of 0.16\,km~s$^{-1}$. The wider bandwidths were necessary to allow the SiO(2--1) and \hh13cop(1--0) lines to be observed together in the 160\,MHz bandwidth and to enable the broad, multiple-$K$ \mecn(5--4) line profile to fit within the 120\,MHz band.
For the $\sim$230\,GHz CO(2--1) and
\c18o(2--1) lines, which both have very wide line profiles, we used the 1~MHz
backend, which offers a bandwidth of 512\,MHz and a spectral resolution of 1.5\,km~s$^{-1}$. Table \ref{ta:freq} gives an overview of the bandwidth and spectral resolution for each transition.
We observed in position-switching mode, where the off position\footnote{The off position is an observation of `blank' sky, which is observed to remove the
instrumental bandpass structure.} was located 800\arcsec\ away.

During the observations, we performed a pointing check every 1.5 hours on a nearby
quasar or on the H{\sc ii}\,regions G10.6--0.4 or G34.3+0.2. The pointing was found
to be accurate within 4\arcsec. The focus check was usually performed on Jupiter or 3C\,273 at the beginning of each
observing run. The opacity at 230\,GHz was variable from excellent winter weather conditions
($\tau=0.1$) to average summer conditions of $\tau=0.5$. The system temperatures at 230\,GHz ranged from 270--920\,K.
At 100\,GHz the opacity ranged from 0.04--0.1, and the system temperatures
were between 87 and 173\,K.

The observed output counts were calibrated to antenna temperatures, $T^\star_{\mathrm{A}}$, using the standard chopper-wheel technique (\citealt{kutner:1981})
The  $T^\star_{\mathrm{A}}$
temperature is the brightness temperature of an equivalent source that fills
the entire $2\pi$ radians of the forward beam pattern of the telescope. This
can be converted to a main beam brightness temperature, $T_{\mathrm{MB}}$, by multiplying by the ratio of the forward efficiency, $F_{\mathrm{eff}}$, and the main beam efficiency, $B_{\mathrm{eff}}$:
\begin{equation}
T_{\mathrm{MB}} = \frac{F_{\mathrm{eff}}}{B_{\mathrm{eff}}} \times T^\star_{\mathrm{A}}\,.
\end{equation}
Table \ref{ta:freq} lists the efficiencies for each transition.
\subsection{APEX observations}

The observations were carried out with the APEX 12m submillimeter telescope located on the Chajnantor plateau in the Atacama desert (Chile) during several runs between 2007 June 10 and November 2.
 We used the double sideband receiver
APEX-2A equipped with two fast Fourier transform spectrometer (FFTS) backends (\citealt{risacher:2006,klein:2006}). The signal and image sidebands are separated by 12\,GHz. We used two
setups in our observations, which are described in Table \ref{ta:freq}. Each
FFTS has a bandwidth of 1 GHz, and 8192 channels, which corresponds to
a velocity resolution of $\sim$0.12--0.15\,km~s$^{-1}$ for lines between
285--350\,GHz. For the \h2co(4--3) rotational transition, we observed only the $4_{04}-3_{03}$ $K$ level, since the remaining $K$ levels of this transition were outside the bandpass due to a mistake in the center frequency.

Each observation was performed with on-off iterations using three subscans, integrating in total between 90--230 seconds on source.
From the IRAM 30m observations, which were carried out before the start of our APEX runs, it was clear that the off positions were often contaminated by extended CO emission. 
Therefore, we used for the APEX observations off positions much further out, at 1800\arcsec , which in most cases were free of emission.

The calibration of the APEX data was, just as with the IRAM 30m telescope, carried out using the chopper-wheel technique. The telescope efficiencies, necessary to calculate the main beam brightness temperature, are listed in Table\.ref{ta:freq}. Before each run, the focus was adjusted on a planet, usually Jupiter. Pointings were made once every three hours on nearby H{\sc ii}\,regions with strong dust continuum emission, such as Sgr\,B2(N). The precipitable water vapor was between 1.44 and 2.80\,mm, while the system temperatures ranged from 190--300\,K at 280\,GHz and 410--575\,K for 356\,GHz.

\subsection{Data reduction}

The processing of the data, such as smoothing (spectral averaging) to increase the signal-to-noise ratio and baseline subtraction, was done in CLASS (\citealt{pety:2005}).
For most lines, Gaussian fitting was performed to retrieve the basic line
parameters, such as line widths, $v_\mathrm{LSR}$, and line intensities. For the lines with hyperfine structure (hfs), we used the hfs method, which allows additionally the derivation of the optical depth of the main hfs component (when applicable). The spectral plots were also prepared in CLASS.

\begin{table*}
\centering
\caption{Molecular lines and frequencies\label{ta:freq}}
\begin{tabular}{lcrrrrccccc}
\noalign{\smallskip}
\hline
\hline
\noalign{\smallskip}
Molecule   & $J+1\rightarrow J$ & $E_u/k_B$&Frequency & $n_\mathrm{crit}$\tablefootmark{a} & Bandwidth & Resolution &Beam &$B_\mathrm{eff}$&$F_\mathrm{eff}$& 1$\sigma$
r.m.s\\ 
           &                & (K)&(GHz)   & (cm$^{-3}$)& (MHz) & (km~s$^{-1}$) & (\arcsec) & && (K)  \\ 
\noalign{\smallskip}
\hline
\noalign{\smallskip}
\multicolumn{10}{c}{IRAM 30m telescope}\\
\noalign{\smallskip}
H$^{13}$CO$^+$ & 1--0 &  4.16~~      & 86.754~~ & $1.7\times10^5$  & 160 & 0.16& 29 &0.78 & 0.98 & 0.06     \\ 
SiO                          & 2--1   &  6.25~~    & 86.847~~  & $7.3\times10^5$ & 160 & 0.16& 29 &0.78 & 0.98  &0.07       \\
HCO$^+$              & 1--0   &  4.28~~     & 89.189~~  & $1.8\times10^5$ & 40 & 0.08 &28 & 0.78 & 0.98  &0.10       \\ 
CH$_3$CN           & 5--4, $K$=0, 1, 2, 3   &  13.24\tablefootmark{b}      & 91.987\tablefootmark{b}  & $4.7\times10^5$ & 120 & 0.16&27  & 0.78 & 0.98 &0.09       \\ 
N$_2$H$^+$       & 1--0     &  4.47~~     & 93.174~~ & $1.6\times10^5$  & 40 & 0.08&27 & 0.78 & 0.98 &0.11        \\ 
C$^{18}$O            & 2--1    &    15.81~~   &  219.560~~  & $9.2\times10^3$ & 512 & 1.5&12  & 0.62 & 0.94 &0.12   \\
CO                         & 2--1      &   16.60~~   & 230.538~~ & $1.1\times10^4$ & 512 & 1.5&11 & 0.58 & 0.92  &1.3        \\
\noalign{\smallskip}
\multicolumn{10}{c}{APEX telescope}\\
\noalign{\smallskip}
N$_2$H$^+$ & 3--2   &   26.83~~   & 279.512~~ &$3.0\times10^6$ & 1000&  0.15 & 22&0.73 &0.97&0.23 \\
H$_2$CO    & $4_{04}$--$3_{03}$\tablefootmark{c} &34.90~~ & 290.623~~ &$9.2\times10^6$ & 1000&  0.15 & 22&0.73 &0.97&0.20  \\
\hcop\     & 4--3     &    42.80~~   & 356.734~~ &$9.1\times10^6$ & 1000&  0.12 & 18&0.73 &0.97&0.47\\
CO         & 3--2       &   33.19~~  & 345.796~~ &$3.5\times10^4$ & 1000&  0.12 & 18&0.73 &0.97&1.1\\
\noalign{\smallskip}
\hline
\hline
\end{tabular}
\tablefoot{Columns are (from left to right) the molecule, its transition,  its upper energy level, its frequency, its critical column density, the bandwidth used in the observation, the velocity resolution of the observation, the telescope beam, the main beam and forward beam efficiency of the observation, and the mean 1$\sigma$ noise value.
\tablefoottext{a}{Calculated from the collision rates at $T=20$\,K from the LAMBDA molecular database (\citealt{schoier:2005}).}
\tablefoottext{b}{Frequency and $E_u/k_B$ for the $K$=0 level. Observed were $K$=0, 1, 2, 3, 4}
\tablefoottext{c}{The observed $K$ level is given in subscript.}
}
\end{table*}

\section{Results}

An overview of all the detected lines per clump and the clump J2000 positions are given in Table \ref{ta:det}. For sources with no detection, we put a 3$\sigma$ upper limit (also in Table~\ref{ta:det}).
\subsection{Infall}
\label{sect:infall}

The infall signature of a  source can be recognized by a line profile with a double-peaked structure, where the intensity of the blue peak exceeds the intensity of the red peak
(\citealt{leung:1977,zhou:1993,myers:1996,mardones:1997, evans:1999}). 
Infall models assume a source where the infall velocity, density, and excitation temperature increase toward the center. In this scenario, the 
blue peak of an optically thick self-absorbed line originates in the rear part, and the red peak in the front part of the infalling shell. 
An increase in the infall velocity,  $v_\mathrm{in}$, can cause the red peak to
diminish; at very large values of $v_\mathrm{in}$, the red peak can even
disappear and become a red ``shoulder''.
A red excess, in contrast to this blue excess, may be caused by expansion or
outflow. Alternatively, the red excess can be caused by an outwards moving blob of
matter, instead of large-scale outward motion (\citealt{evans:1999}). To interpret the infall profile of an optically thick line, it is also necessary to compare it with an optically thin line measured toward the same position. The latter, often a line from a rare isotopologue, shows a maximum, defining the systemic velocity at the self-absorption minimum of the optically thick line.
The observed line profiles can, in addition to large scale motions such as infall, also be influenced by abundance changes though the clouds. Hence, it is desired to investigate infall using several molecules. To probe different depths in the clouds and thereby possible different infall velocities, we used lines with different critical densities: the CO(2--1), CO(3--2), \hcop(1--0), and \hcop(4--3) transitions. 

\subsubsection{CO line profiles}
We started the study with the CO(2--1) and CO(3--2) lines, whose critical
densities are around 1$\times10^4$ and 4$\times10^4\,\mathrm{cm^{-3}}$ (Table \ref{ta:freq}), respectively. 
Since the CO molecule, as a common component of the
interstellar medium (ISM); has a very extended distribution, the off position is not necessarily free of its emission. This can result in
artificial absorption lines in the spectrum. 
To avoid this, one can observe with two different off positions and/or use
different CO transitions. Since higher $J$ transitions have a higher critical
density, their emission is confined to a more compact and higher density region than emission from lower $J$ lines. We employed this strategy and
observed, in addition to the CO(2--1) line with the IRAM 30m, the CO(3--2) line with APEX using different
off positions. Also, because of the molecules' ubiquity and relatively low critical density, CO observations
pick up emission from various Galactic arms. This manifests itself in the
spectrum by more than one peak at various local standard of rest (LSR) velocities. 

We searched for infall signatures by comparing the velocities of the CO(2--1) and CO(3--2) emission peaks with those of the optically thin \c18o(2--1) line.
When the peak of the optically thick line of the main isotopologue appeared shifted blueward of the \c18o\ peak, the source was marked a blue excess source or a blue source. A source with a red shifted peak was marked a red source.

\begin{figure}
\centering
\includegraphics[angle=-90,width=6cm]{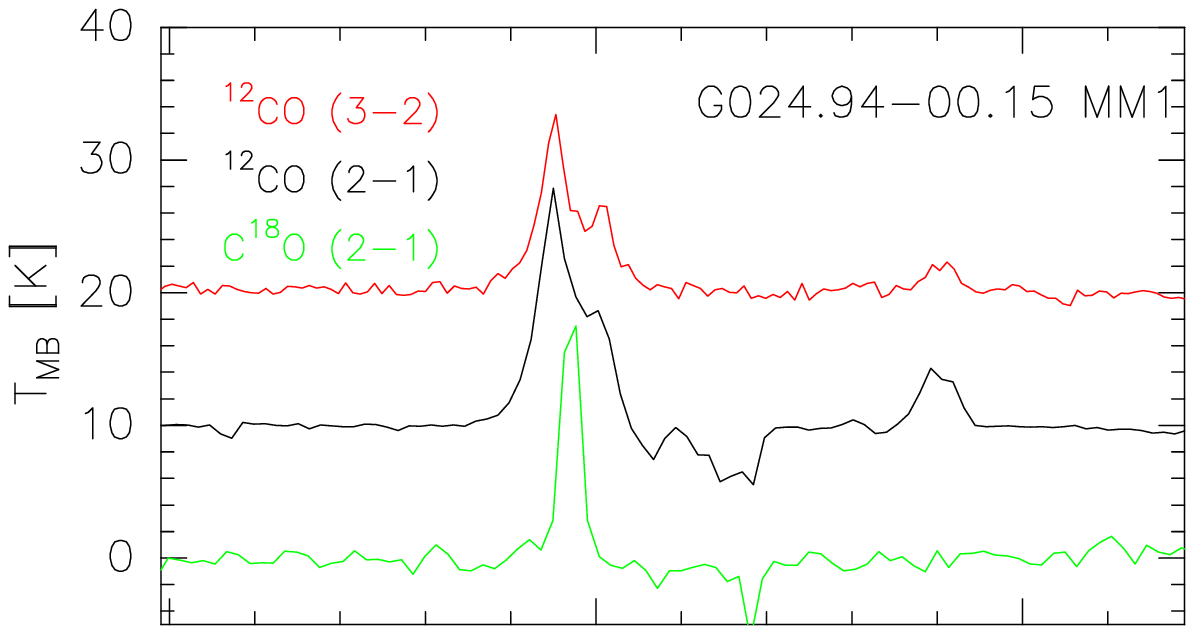}
\includegraphics[angle=-90,width=6cm]{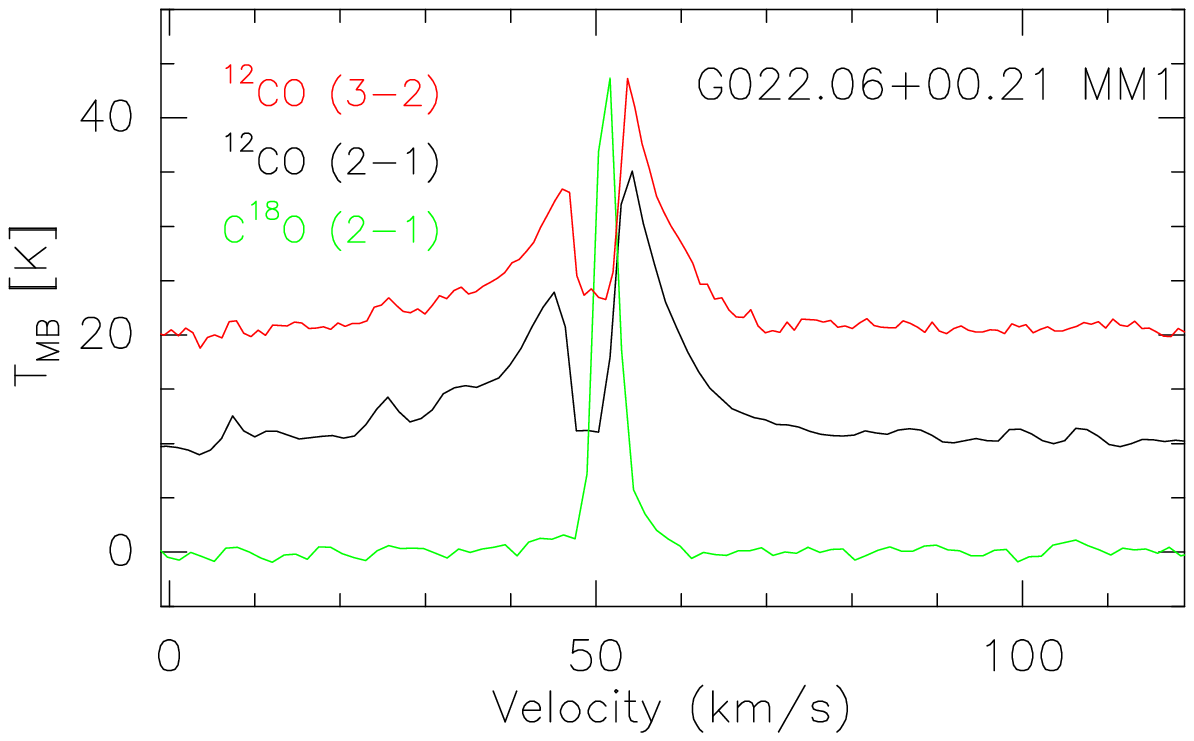}
\caption{Two examples of infall/outflow signatures of the $^{12}$CO line profiles. The optically thin \c18o(2--1) emission (scaled for better visibility, shown in green) indicates the systemic velocity. The CO(2--1) and CO(3--2) emission is  shown in black and red, respectively. The panels show infall ({\sl top }) and outflow ({\sl bottom}).\label{fig:co}}
\end{figure}

In general, both the CO transitions showed similar behavior (see Table \ref{ta:infall}).
Two examples of infall and outflow signatures are presented in
Fig. \ref{fig:co}. When the peaks of the optically thick and thin lines
were at the same velocity, nothing could be inferred.
Mostly, the CO lines were broader than the \c18o\ lines and had line wings extending from 10 to almost 50\,$\mathrm{km~s^{-1}}$ relative to the systemic LSR velocity. Line width broadening
is partially due to a high optical depth, which is higher for the CO lines than for \c18o\  lines. However, the wings are dominated by gas undergoing strong and dominant motions, such as infall or outflow, and in cases such as G022.06+00.21\,MM1 by high-velocity emission features (`bullets').

The results for CO are given in Table \ref{ta:infall}. For the lower transition, the presence of wings is indicated, and, if present, their velocity range is given.  
The CO line profiles delivered the first indications of infall or outflow. A deeper study was done with the \hcop(1--0) and \hh13cop(1--0) lines, reported in the next section.

\subsubsection{The skewness parameter $\mathbf{\delta v}$}

\begin{figure}
\centering
\includegraphics[angle=-90,width=6cm]{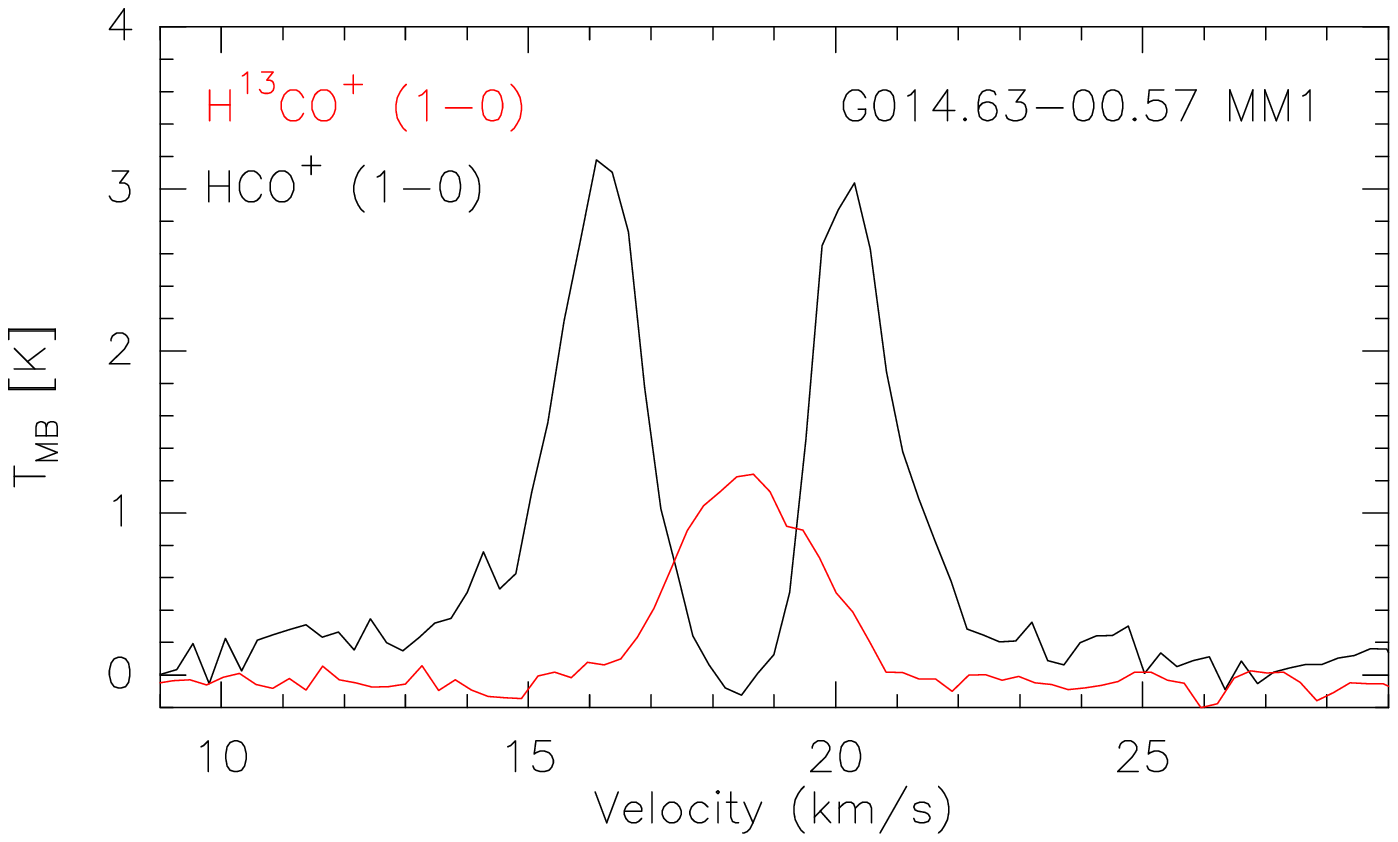}
\includegraphics[angle=-90,width=6cm]{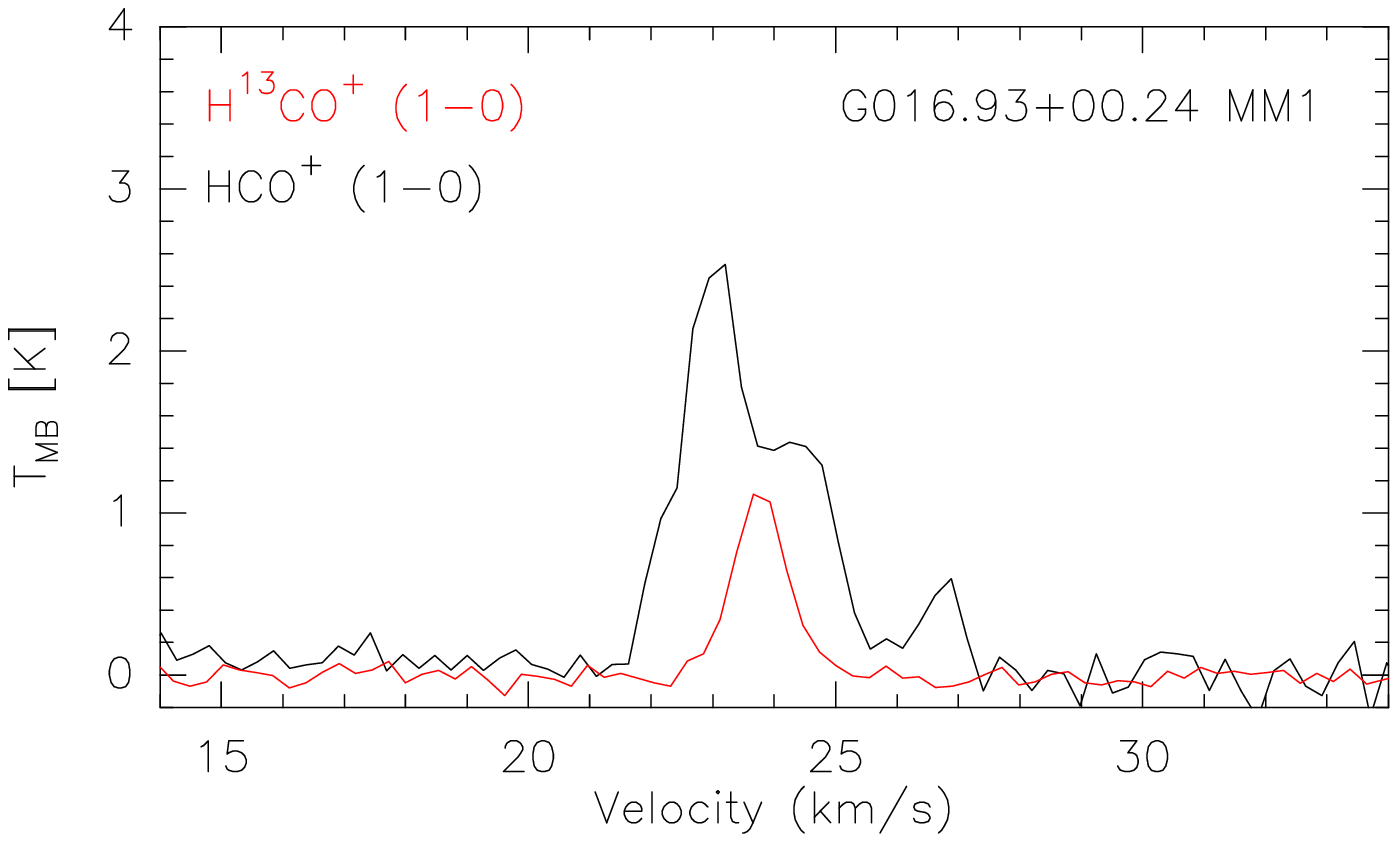}
\includegraphics[angle=-90,width=6cm]{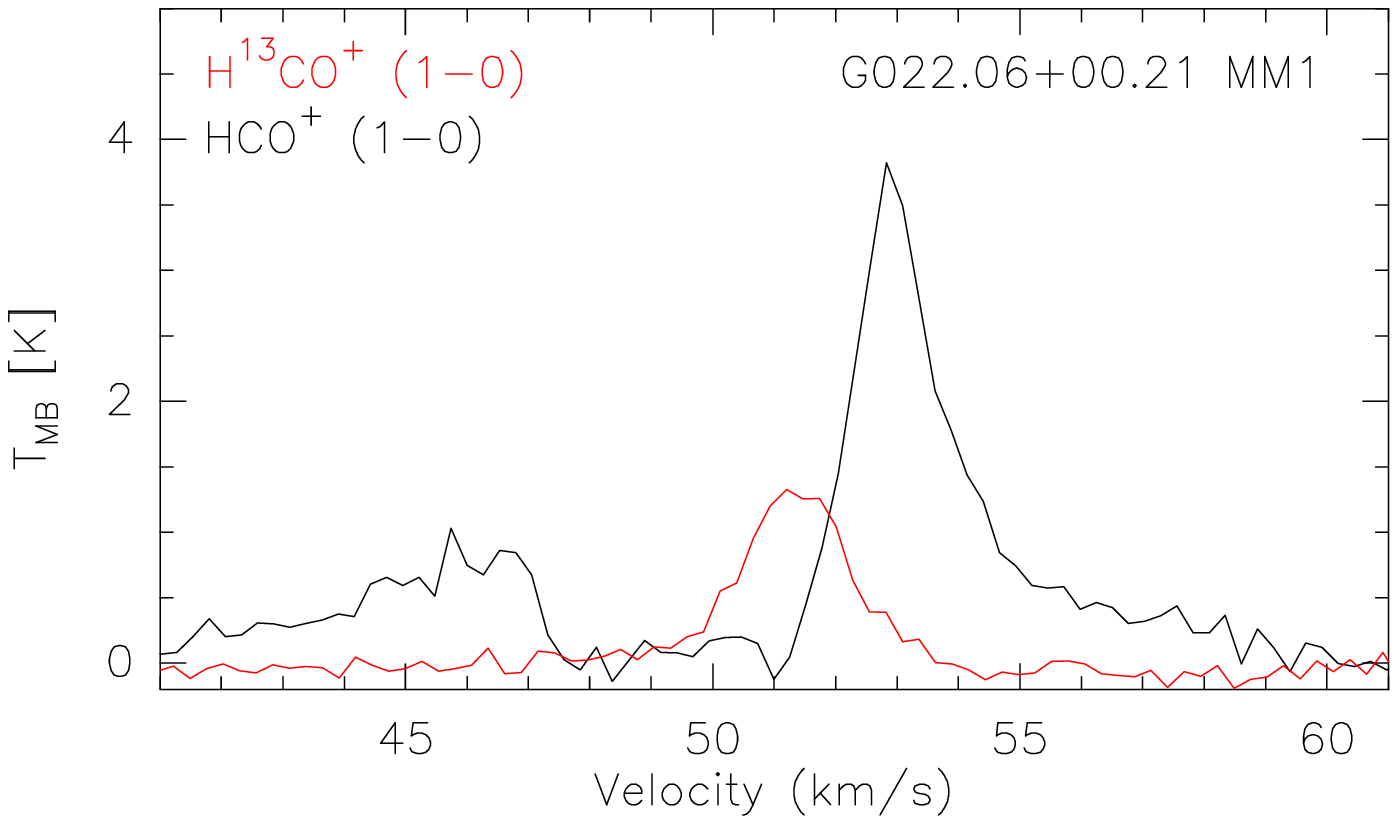}
\caption{Three examples of infall/outflow signatures of the \hcop(1--0) line profile. Red marks both \hh13cop(1--0), which is optically thin, and the systemic velocity of the dense gas. Compared to this the shift of the peak of the \hcop(1--0) emission, in black, becomes clear. The panels show infall ({\sl middle and bottom}), outflow  ({\sl bottom right}), and a case of central self-absorption, where both peaks are equal  ({\sl top}). The profiles show various degrees of self-absorption.\label{fig:hcopdv}}
\end{figure}

The optically thick \hcop(1--0) line profiles were used for an in-depth study of infall and outflow motions in the clumps. The systemic velocity of a clump was determined from
the optically thin \hh13cop(1--0) line. Examples of \hcop(1--0) line profiles are shown in Fig.~\ref{fig:hcopdv}.
For half of the clumps, the \hcop(1--0) line showed a double-peaked profile with a self-absorption dip. Here, the ratio of the two intensity peaks, $\frac{T_{\mathrm{blue}}}{T_{\mathrm{red}}}$, was used to check whether the source showed blue or red excess in the emission (ratio listed in Table \ref{ta:infall}).

When the \hcop(1--0) line profile showed only one peak, we compared its velocity with the velocity of the optically thin \hh13cop(1--0) peak. The measure of this shift in velocity is called the skewness parameter. 
Following the
method outlined in \citet{mardones:1997}, the skewness parameter $\delta v$, is defined as
\begin{equation}
\delta v = \frac{v_\mathrm{thick} - v_\mathrm{thin}}{\Delta v_\mathrm{thin}}\,, 
\end{equation}
where $v_\mathrm{thick}$ and $v_\mathrm{thin}$ are the LSR velocities of the peaks
of the optically thick and thin lines, respectively; $\Delta v_\mathrm{thin}$
is the line width of the optically thin line. The line width and LSR velocity
of the \hh13cop(1--0) line were retrieved by Gaussian fits (Table \ref{ta:hcop}). For the
optically thick \hcop(1--0) line, the profiles often showed non-Gaussian shapes; therefore, the position of the peak was determined by eye (also given in Table \ref{ta:hcop}). After \citet{mardones:1997}, we adopt the definition of significant blue and red excess, namely for the former if  $\delta v \leq -0.25$ and the latter when $\delta
v \geq 0.25$.

The skewness parameter was determined for the \hcop(1--0), \hcop(4--3), and CO(3--2) lines (all listed in Table \ref{ta:infall}). For all molecules, the values ranged between --1.5 and 1.5, except for G14.39-00.75B where the $\delta v$ of the \hcop(1--0) line was --5.79. The latter is due to a very blue shifted \hcop(1--0) peak, most likely caused by a high self-absorption in this source.
The clumps in high extinction clouds have skewness parameters similar to values in the literature: --0.5 for UCH{\sc ii}\,regions
(\citealt{Wyrowski:2006}), --1.5 to 1 for HMPOs (\citealt{fuller:2005}), --1 to
2 for Extended Green Objects (EGOs, \citealt{chen:2009}), which are shocked regions, and --0.5 for the cluster-forming clump G24.4 (\citealt{wu:2005}). 

\begin{figure*}
\centering
\includegraphics[angle=-90, width=11cm]{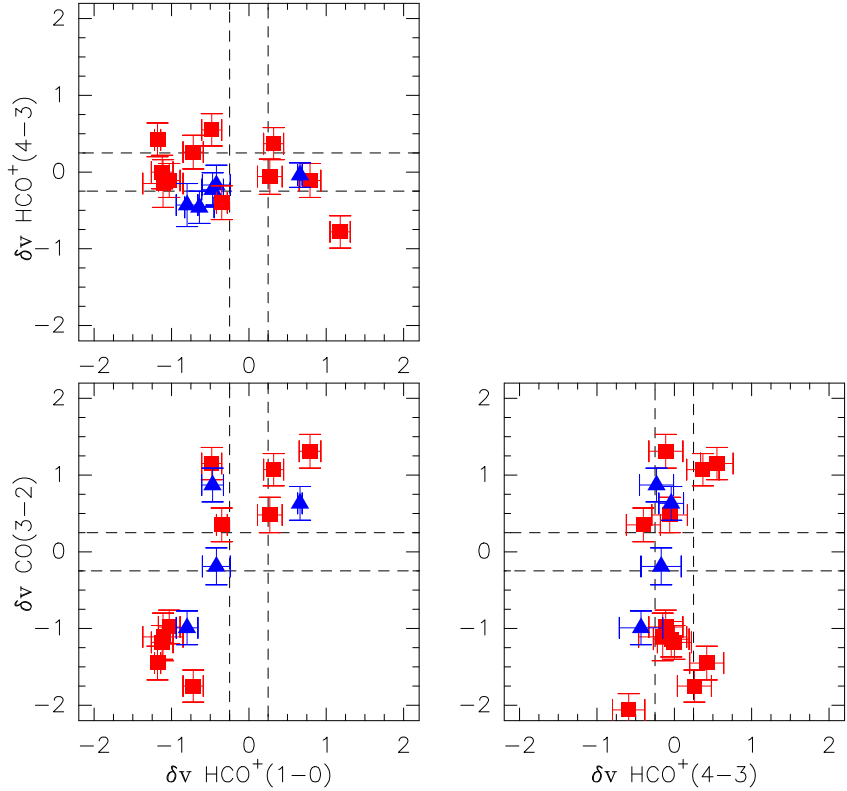}
\caption{\label{fig:infallcomp} Skewness parameter of the \hcop(1--0) line versus that of the
  \hcop(4--3) and $^{12}$CO(3--2) lines. The dashed lines mark the boundary of significant excess at $\delta v >|0.25|$. The blue triangles represent clumps in peaked clouds, while the red squares represent sources in multiply peaked clouds.}
\end{figure*}

Cross correlations of the skewness between the three different molecules (Fig.~\ref{fig:infallcomp}, clumps from peaked and multiply peaked clouds are color coded in light of the evolutionary sequence discussion in Sect.~\ref{sec:evodis}) show that the \hcop(4--3) line is the least sensitive to infall and/or outflow, while the \hcop(1--0) seems the most sensitive. Most of the \hcop(4--3) lines' $\delta v$ values are $<|0.25|$, while those of the \hcop(1--0) and CO(3--2) lines exhibit a range of skewness. Also, \citet{fuller:2005} found that the transitions with higher critical densities show less infall. These observations suggest that the infall occurs predominantly in the low-density environment, which is in contrast with the increasing infall velocity toward the center that is expected in the collapsing clump model. The observations can possibly be explained by the different optical depths of the \hcop\ transitions, because an infall profile requires the transition to be sufficiently opaque. 
Depending on the density of the infall environment, the transitions with a high critical density might not be opaque enough to observe the skewness profile, while transitions of lower critical density will be already sufficiently optically thick to observe the infall profile. For the \hcop(1--0) and CO(3--2) lines, the skewness seems to be correlated, which makes the infall determination based on these two molecules more reliable. 

The excess parameter shows the average behavior of the blue excess over the red excess sources. It is defined as $E = (N_\mathrm{blue}-N_\mathrm{red})/N_{\mathrm{tot}}$, where $N_\mathrm{red}$ is the number of red excess sources, $N_\mathrm{blue}$ the total number of blue excess sources, and $N_\mathrm{tot}$ the total
number of all sources (\citealt{mardones:1997}). 
\begin{table}
\centering
\caption{Distribution of the skewness parameter per molecule\label{ta:bintest}}
\begin{tabular}{lrrrrrrr}
\hline\hline
\noalign{\smallskip}
Transition & $N_\mathrm{blue}$ & $N_\mathrm{red}$ &$N_\mathrm{skew}$& $N_\mathrm{tot}$ & $E$     &$P$  &  $<\delta v>$\\
\noalign{\smallskip}
\hline
\noalign{\smallskip}
\hcop(1--0) & 24               & 14            &38  &  47             &  0.22   & 0.07 & $-0.26$\\ 
\hcop(4--3) & 5                & 4              &9 &  18              &  0.06   & 0.50 & $-0.11$\\
CO(2--1)\tablefootmark{a} & 24           & 11            &35  &  35              & 0.37     & 0.02 & -\\
CO(3--2) & 8             & 7            &15   &  16             & 0.06      & 0.50 & $-0.31$\\
\noalign{\smallskip}
\hline
\end{tabular}
\tablefoot{Columns are (from left to right) number of blue excess sources, number of red excess sources, total number of sources with a significant skewness, total number of sources, the excess parameter, the probability of the distribution to arise by chance, and the average skewness.}
\tablefoottext{a}{The numbers of the CO(2--1) transition serve only as an indication.}
\end{table}

Table \ref{ta:bintest} lists the excess parameter and average skewness parameter by transition. For each transition, we calculated, through the binomial test, the probability $P$ that the distribution between blue and red excess sources occurred by chance. The binomial distribution is defined as  
\begin{equation}
P = \binom {n}{k} p^{k} (1-p)^{(n-k)}\,,
\end{equation}
 where $n$ is the total number of trials, $k$ the number of successes, and $p$ the success probability. In our case, $n$ is the total number of blue and red excess sources, $k$ the number of
 blue excess sources, and the success probability $p$ equals $0.5$ if the blue and red excess sources are randomly distributed. We can then calculate the possibility that the distribution of the number of blue sources {\em equal or higher} than the number observed arises by chance by adding all possibilities $P$($n$, $k$, $p$) + $P$($n$, $k$+1, $p$) + .. until $k=n$. 
 
The small probability $P$ of 7\% for \hcop(1--0) line in Table \ref{ta:bintest} indicates that there are significantly more blue excess sources than expected for a random distribution. On the other hand, the higher CO(3--2) and \hcop(4--3) transitions have a probability of 50\%. Hence, the number of blue excess sources is likely by chance and not significant. The results for the CO(2--1) line, $P=0.02$, should be treated with care, because the same definitions of blue and red sources do not apply. Due to the possible confusion in the line profiles, the CO(2--1) was just classified based on the peak emission with respect to the \c18o(2--1) line without the calculation of $\delta v$. Therefore, the numbers in Tables \ref{ta:infall} and \ref{ta:bintest} have to be taken just as indications.    
In conclusion, the \hcop(1--0) line shows a significant blue excess and is the best indicator of infall for our sample of clumps. The excess parameter of the \hcop(1--0) line, 0.22, is similar to values found in a previous survey by \citet{fuller:2005} of high-mass protostellar objects reaching excesses of 0.29 and 0.31. Studies of low-mass star formation also report similar numbers (\citealt{mardones:1997, evans:2003}).

\subsection{Temperature and density estimations from ${N_2H^+}$(1--0) and (3--2) transitions}

\begin{figure}
\centering
\includegraphics[angle=-90,width=6cm]{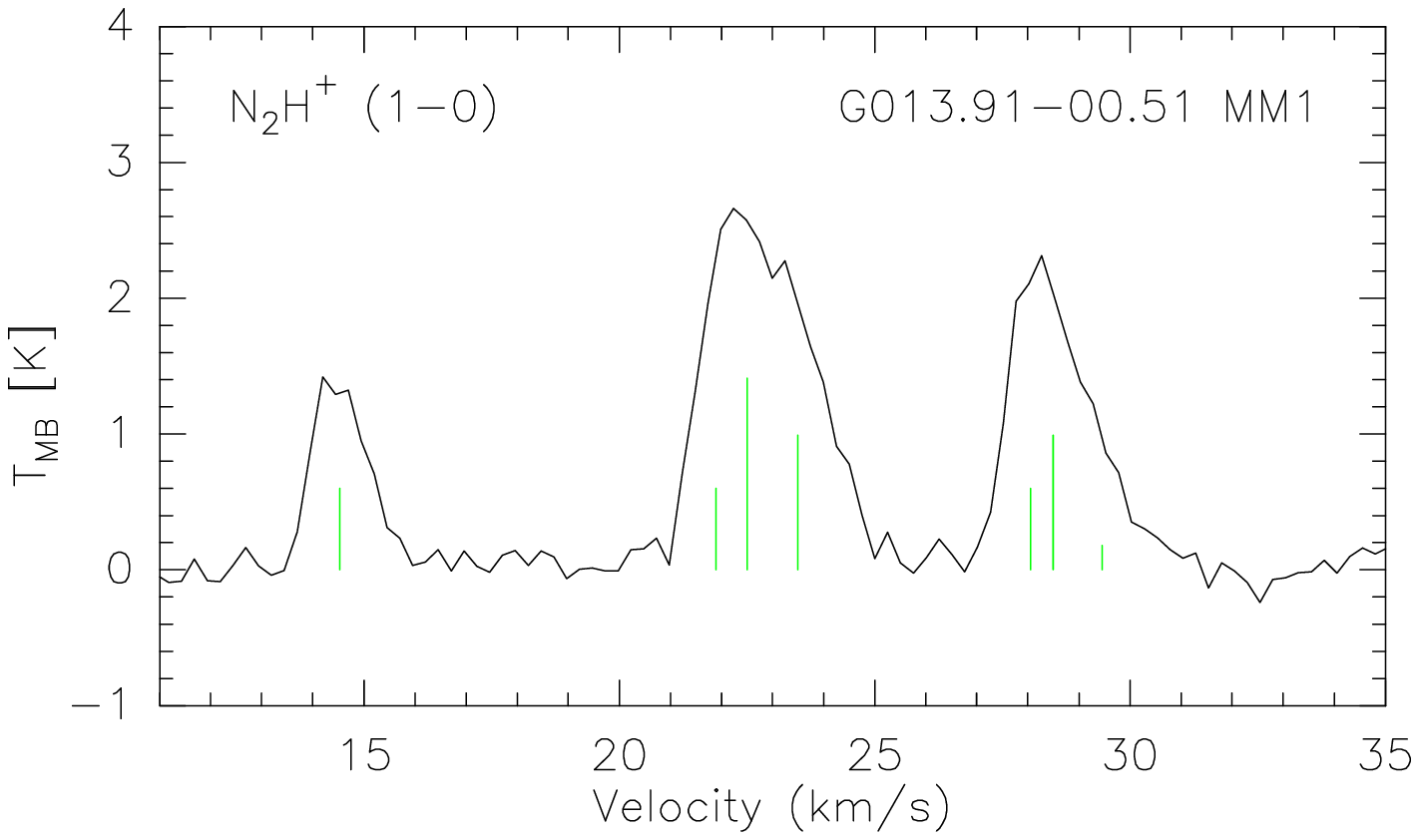}
\includegraphics[angle=-90,width=6cm]{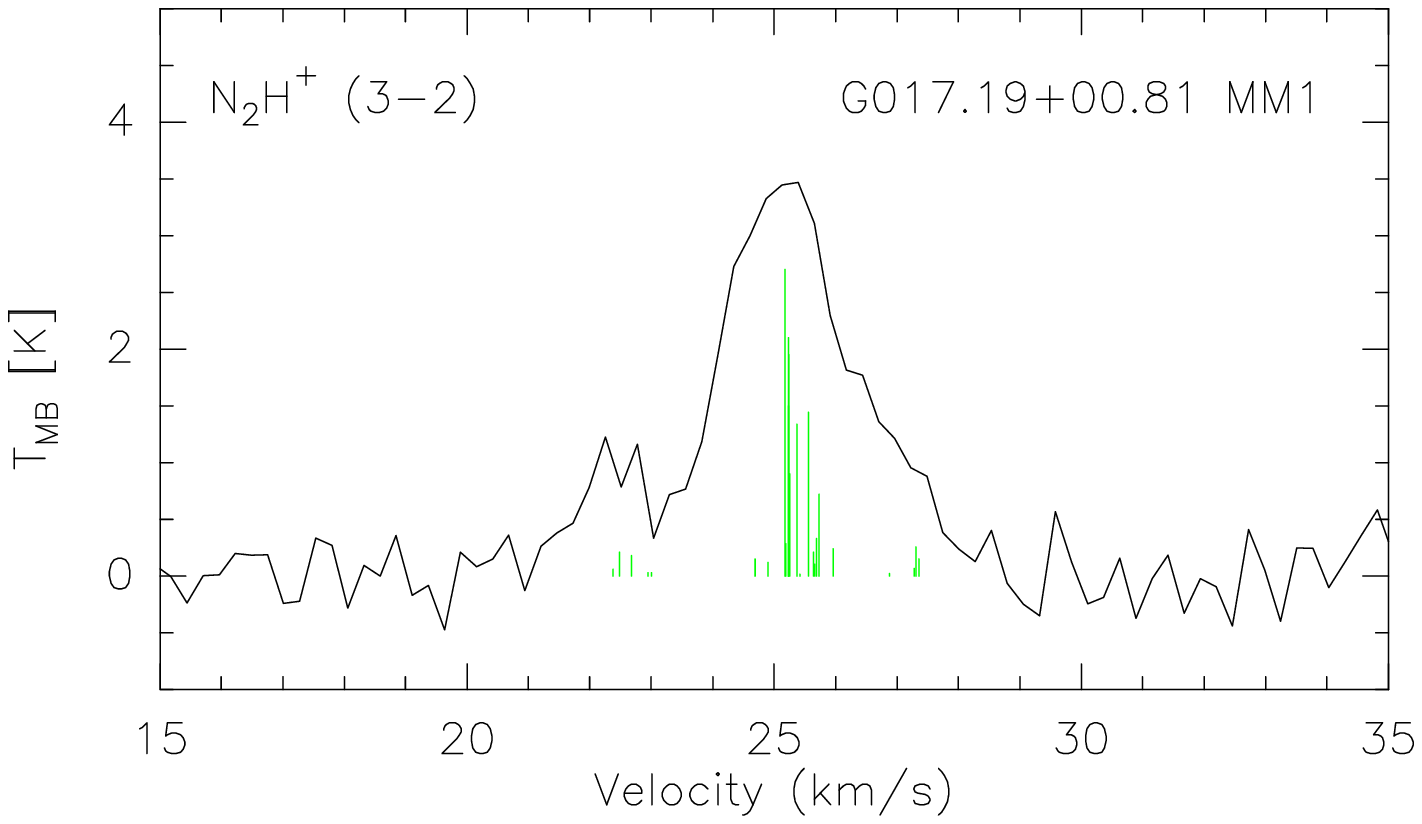}
\caption{\label{fig:n2hp} \n2hp(1--0) ({\em top}) and  \n2hp(3--2) ({\em bottom}) line profiles. The green lines show the locations of the hyperfine components.}
\end{figure}

The \n2hp(1--0) and \n2hp(3--2) rotational transitions have hfs arising from the interaction between the molecular electric field gradient and
the electric quadrupole moments of the two nitrogen nuclei
(\citealt{caselli:1995}). The \n2hp(1--0) rotational transition is split into
seven hyperfine components. The three main hyperfine groups were well resolved
in our observations (see Fig. \ref{fig:n2hp}), and for a few sources we could fit all seven components. In the higher transition, we were barely able to even resolve the three main groups of hfs components (Fig. \ref{fig:n2hp}).

For fitting the \n2hp(1--0) transition we took the hfs into
account by using the hfs method for the line fitting in CLASS \footnote{http://www.iram.es/IRAMES/otherDocuments/postscripts/classHFS.ps}. This method assumes one excitation temperature and one line width for all seven hyperfine components, as well as the fact that the opacity as a function of frequency for each hfs has a gaussian shape. Then, besides the intrinsic line width and integrated intensities, the hfs method also determines the optical depth of the main hfs group. The total optical depth $\tau_\mathrm{tot}$ can be calculated from the obtained main hfs group optical depth $\tau_\mathrm{main}$ by multiplying it by the sum of relative intensities of the satellites $S$ as $\tau_\mathrm{tot}=S\tau_\mathrm{main}$. In case the relative intensities are scaled so that their sum equals unity, the hfs fit directly delivers the total optical depth (the sum of all the hfs optical depths).
We calculated the column density of \n2hp, $N_\mathrm{N_2H^+}$ (\ref{eq:n2hp})  following \citet{benson:1998}.
The results of the \n2hp(1--0) line profile fitting; the integrated intensity (including the hyperfine components), total optical depth, and line width are placed together with the
excitation temperature and the column density in Table \ref{ta:n2hp10}. For \n2hp(3--2), Table \ref{ta:apex1} lists the parameters obtained by Gaussian fits. We did not observe the hfs as clearly as in the lower transition and could therefore not determine the optical depth for the higher transition.

For sources with $\tau_\mathrm{(1-0),\,tot}\lesssim0.4$, which usually also had high relative errors, we found unrealistically high $T_\mathrm{ex}$ values that were much higher than the ammonia rotational temperatures, $T_\mathrm{rot}$ (Paper I). The CLASS hfs method does not provide good estimates for excitation temperatures (hence also column densities) when the relative errors in optical depth are very large ($>40$\%) and when the optical depth is very small ($\tau_\mathrm{(1-0),\,tot}\lesssim0.4$) (see, e.g., \citealt{fontani:2006,fontani:2011}). 
After discarding these sources, the excitation temperatures generally ranged from 3 to 29\,K, averaging 8.1\,K . In fact, only two sources had a $T_\mathrm{ex}>20\,K$; G017.19+00.81 MM2 (28\,K) and MM3 (29\,K). We assumed that the \n2hp\ gas has a similar kinetic temperature as the \amm\ gas and that they are both in local thermal equilibrium (LTE), in which case the $T_\mathrm{ex}$ should equal $T_\mathrm{rot}$. However, the determined excitation temperatures were, with the exception of G017.19+00.81 MM2 and MM3, all lower than the rotational temperature from ammonia. The difference between the $T_\mathrm{ex}$ determined here and the $T_\mathrm{rot}$ determined by
ammonia is the \n2hp\ filling factor. While $T_\mathrm{ex}$, is inversely proportional to the filling factor (see \ref{app:n2hp}),
the rotational temperature is an intrinsic temperature (independent of
$f$ since it is solely determined by line ratios). We estimated the filling factor from the ratio of the excitation temperature of \n2hp(1--0) and the \amm\ rotational temperature:
\begin{equation}
f_{\mathrm{N_2H^+(1-0)}}\sim\frac{T_\mathrm{ex,\,N_2H^+}}{T_\mathrm{rot,\,NH_3}}.
\end{equation}
The individual filling factors range from 0.2 to 0.8. For clumps G17.19+00.81 MM2, and MM3, we found meaningless filling factors larger than unity, which is possibly due to an overestimation of the excitation temperature. A filling factor smaller than unity indicates either that the source size is smaller than
the beam or that the source is clumpy instead of centrally condensed. Alternatively, the \n2hp(1--0) line could be sub-thermally excited, a consequence of densities lower than the critical density. At such densities, a transition's upper energy level becomes underpopulated: collisions do not manage to populate it, and the system is in non-LTE. In this case, the excitation temperature will be lower than the kinetic temperature, thus mimicking the behavior of a small filling factor. 

\begin{figure}[!h]
\centering
\includegraphics[angle=-90,width=7cm]{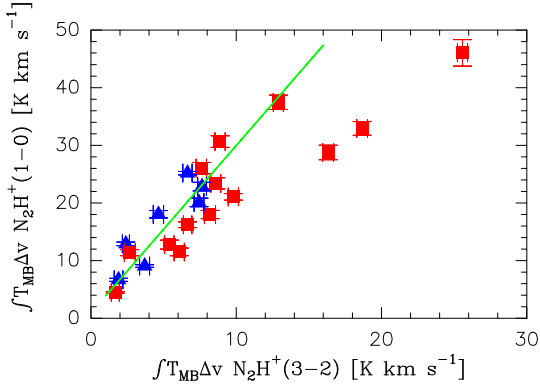}
\caption{\label{fig:n2hp-ratio} \n2hp(1--0) integrated intensity versus \n2hp(3--2) integrated intensity for all detected sources. The blue triangles represent clumps in peaked clouds, while the red squares represent sources in multiply peaked clouds. The green solid line marks the trend of the observed ratio of 2.9.}
\end{figure}

To have another estimate of the filling factor, we performed radiative transfer modeling to obtain the intrinsic \n2hp\ column density. In a non-LTE system, the ratios of line intensities deviate from the LTE predictions and become a function of density, column density, temperature, and optical depth. In particular, we consider the \n2hp(1--0)/\n2hp(3--2) line ratio and combine this with non-LTE radiative transfer modeling using RADEX (\citealt{tak:2007}) and the molecular database LAMBDA (\citealt{schoier:2005}). 
Figure \ref{fig:n2hp-ratio} shows the integrated intensities of the two \n2hp\ transitions plotted against each other. We use differently colored symbols
for each cloud class (diffuse, peaked, and multiply peaked) in the graphs shown throughout this paper to visualize the (possible) behavior for different phases of cloud evolution.
The integrated intensity, $\int T_{\mathrm{MB}}\mathrm{d}v$, of the lower transition, \n2hp(1--0), exceeded that of \n2hp(3--2) by an average of
\begin{equation}
<\frac{\int T_{\mathrm{MB}}\mathrm{d}v(\mathrm{N_2H^+}(1-0))}{\int T_{\mathrm{MB}}\mathrm{d}v(\mathrm{N_2H^+}(3-2))}>=2.9.\\
\end{equation}
In this ratio, we assumed that the filling factor ratio is 1, since it is likely that the filling factors of the two transitions cancel out as the APEX beam is only slight smaller than the IRAM 30m  beam (Table \ref{ta:freq} lists all beamsizes). There are three sources that lie far from the average: G014.63--00.57 MM1, G017.19+00.81 MM2, and G022.06+00.21 MM1, which are all bright at 24\,$\mu$m and contain water masers (see Table \ref{ta:det}).

\begin{figure*}[!h]
\centering
\includegraphics[angle=-90,width=13cm]{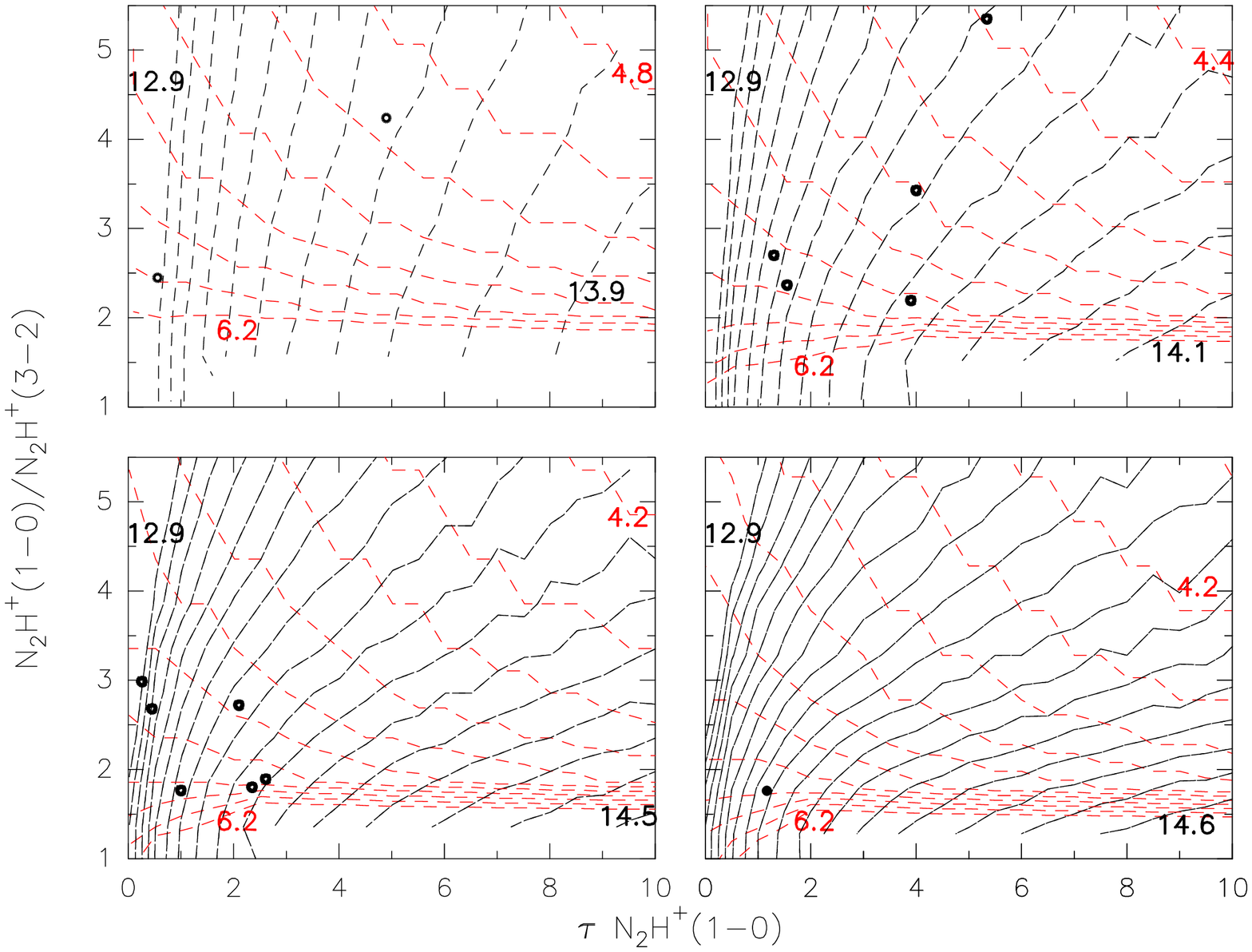}
\caption{\label{fig:Nn-n2hp} \n2hp(1--0) optical depth versus the ratio of \n2hp(1--0) and \n2hp(3--2) integrated intensities. The four plots represent RADEX calculations with different temperatures: top left 10\,K; top right 15\,K; bottom left 20\,K; bottom right 25\,K. The black dashed contours represent the logarithm of the line width times the column density, $\log(\mathrm{N_{N_2H^+}}/ \Delta v )$ with a contour step of 0.1. The red dashed contours are $\log(n_{\mathrm{H_2}})$ with a contour step of 0.2. The lowest and highest contour values are given in each plot. The values of the sources in Table~\ref{ta:RADEX} are plotted by black dots.}
 \end{figure*}

The observed \n2hp\ column density is roughly proportional (depending on the value of $T_\mathrm{ex}$, see Eq.~\ref{eq:n2hp}) with excitation temperature, $N_\mathrm{N_2H^+} \propto T_\mathrm{ex}^{1.4}$ for 5\,K$<T_\mathrm{ex}<$10\,K, which in turn is inversely proportional with filling factor. Hence, the ratio of observed \n2hp\ column density (from the \n2hp(1--0) line) and the intrinsic column density calculated by RADEX (based on the \n2hp(1--0)/\n2hp(3--2) integrated intensity ratio) can be interpreted as a result of the filling factor of the \n2hp(1--0) line. 
Using RADEX, we estimated the behavior of $N_\mathrm{N_2H^+}/\Delta v$ and $n_\mathrm{H_2}$ for different values of the \n2hp(1--0)/\n2hp(3--2) line ratio and the opacity of \n2hp(1--0) for kinetic temperatures of 10, 15, 20, 25\,K. We chose this temperature range based on rotational temperatures derived from the ammonia observations (see also Table \ref{ta:sample}). The results are shown in Fig.~\ref{fig:Nn-n2hp} and Table \ref{ta:RADEX}.

\begin{table*}
\centering
\caption{Results of RADEX calculations based on the (1--0)/(3--2) \n2hp\ ratio \label{ta:RADEX}}
\begin{tabular}{llccccccccc}
\hline
\noalign{\smallskip}
 &&\multicolumn{3}{c}{\underline{\hspace{1.2cm}\n2hp(1--0)\hspace{1.2cm}}} & \underline{\hspace{0.0cm}\n2hp(3--2)\hspace{0.0cm}}& &&&&\\
Source  name& &$\int T_{\mathrm{MB}}\mathrm{d}v$&$\Delta v$ &$\tau$&$\int T_{\mathrm{MB}}\mathrm{d}v$&$T_\mathrm{rot}$ &$N_{\mathrm{N_2H^+, obs}}$ &$N_{\mathrm{N_2H^+, R}}$&$n_\mathrm{N_{H_2},obs}$&$n_\mathrm{N_{H_2},R}$\\
 & & (K~km~s$^{-1}$) & (km~s$^{-1}$) & & (K~km~s$^{-1}$) & (K) & ($10^{13}\mathrm{cm^{-2}}$) & ($10^{13}\mathrm{cm^{-2}}$) &($10^5\mathrm{cm^{-3}}$) &($10^5\mathrm{cm^{-3}}$)\\  
 \noalign{\smallskip}
 \hline
 \noalign{\smallskip}
G013.91--00.51&MM1 & 12.9 & 1.19 &5.34&  2.4 & 14.1 & 2.2&  5.8 & 2.2&  0.3\\
G014.63--00.57& MM1 & 45.8 &2.46& 2.35 &  25.6 & 18.1 & 8.9 &18.7 & 7.7&3.4\\
G014.63--00.57& MM2 & 17.9 & 1.27 &3.90 & 8.2 & 15.7 & 2.8 & 9.4 & 7.3 & 3.4\\
G017.19+00.81&MM2 &32.9  & 2.21 &1.0    &18.7 &18.7 & 13.5 & 7.7 & 6.3 & 4.0\\
G017.19+00.81&MM3 &25.1  & 1.74 &0.91   & 1.7 & 20.1 & 10.9 &2.7  & 0.8 & 1.8\\
G018.26--00.24& MM1&23.5 & 2.50 & 2.10 & 8.6 & 18.2 & 3.7 & 10.5 & 1.8 & 1.4\\
G018.26--00.24& MM2&26.0 &  2.16& 4.00  & 7.6 & 17.4 & 4.3 & 10.1 & 1.3 & 1.0\\
G022.06+00.21&MM1&28.7 & 2.50 &1.17 & 16.3 & 24.7 & 7.7 & 13.0 & 6.8 & 5.2\\
G024.37--00.15& MM2 & 12.7 & 2.28 &1.55 &5.4 & 15.7 & 2.1 & 6.2 & 1.2 & 2.9\\
G024.94-00.15&MM1 & 20.0 & 2.48 &1.30 & 7.4 & 15.2 & 4.2 & 5.0 & 2.0 & 2.9\\
G030.90+00.00A&MM1 & 22.8 &2.38 & 0.26 & 7.7 & 18.6 & -- & 3.1 & 0.7 & 1.8\\
G034.71--00.63&MM1 & 11.7 & 2.31 & 2.61&6.1 & 17.8 & 1.9 & 17.6 & 1.0 & 3.4\\
G034.71--00.63&MM2 & 11.5 & 1.58 & 4.90& 2.7 & 12.4 &2.1 & 6.3 & 1.1 & 1.4\\
G053.81--00.00 & MM1 & 9.0 &   1.64& 0.56&3.7 & 12.4 & 1.0 & 12.1 & 3.6 & 8.6\\
\noalign{\smallskip}
\hline
\end{tabular}
\tablefoot{Column are (from left to right) source name, integrated \n2hp(1--0) line intensity, FWHP \n2hp(1--0) line width, \n2hp(1--0) optical depth, integrated \n2hp(3--2) line intensity, \amm\ rotational temperature (Paper I), observed \n2hp\ column density, calculated \n2hp\ column density from RADEX, observed hydrogen density from the 1.2\,mm continuum (Paper I), calculated hydrogen density from RADEX. }
\end{table*}

In Fig.~\ref{fig:Nn-n2hp} we marked the positions of the sources from Table \ref{ta:RADEX} at the intersection of the ratio of the (1--0) and (3--2) \n2hp\ intensities and the \n2hp(1--0) optical depth in the plot calculated with a kinetic temperature closest to the \amm\ rotational temperature of the clump. To obtain the intrinsic \n2hp\ column density, we multiplied $N_\mathrm{N_2H^+}/ \Delta v$ by the observed \n2hp(1-0) line width. In the plots, we see that a rising $T_\mathrm{kin}$ increases the column density for a given optical depth and a line ratio, while the volume density decreases. 
For the handful of sources  that we have measured both \n2hp\ transitions, we compared the filling factor based on the column density with the one determined from the temperature. On average, the filling factors agreed within 30\%. Here we excluded clumps G017.19+00.81 MM2 and MM3, since column density-derived filling factor came out larger than unity, and which confirms that for these sources the derivation of the \n2hp\ parameters, from which the excitation temperature and column density are calculated, gave unrealistic values.

The RADEX results for the volume density were almost all within a factor two or less of the measured volume density from the 1.2\,mm continuum (Table \ref{ta:RADEX}). Given the uncertainties in the derivations of both volume densities, we can say that they are in agreement for most of the sources. The \n2hp\ ratio and the \n2hp(1--0) optical depth can therefore be used to estimate the hydrogen volume density and intrinsic \n2hp\ column density of a clump for a given kinetic temperature. 

We estimated the abundance, $\chi$, of \n2hp\ by comparing the $N_\mathrm{N_2H^+}$ corrected by the filling factor from the temperature ratio (for each source individually) to the hydrogen column density, $N_\mathrm{H_2}$, measured from the 1.2\,mm continuum (Paper I): 
\begin{equation}
\chi(\mathrm{N_2H^+})=\frac{N_\mathrm{N_2H^+}}{N_\mathrm{H_2}}. 
\end{equation}
The individual clump abundances are listed in Table \ref{ta:n2hp10}, and range from $0.26-4.0 \times10^{-9}$, with a mean of $1.3\times10^{-9}$.
The true abundances are likely to be slightly lower, since the $N_\mathrm{H_2}$ is a beam-averaged column density and contains an unknown filling factor. Taking this into account, the abundance is in agreement with the value of \citet{pirogov:2007} for high-mass star-forming cores and the results of \citet{ragan:2006, vasyunina:2011} for IRDCs.

\subsection{Presence of young stellar objects}
\subsubsection{Hot molecular cores: $\mathrm{CH_3CN}$ and $\mathrm{H_2CO}$ }
\label{sect:myso}
\mecn\ and \h2co\ are usually found toward
YSOs or hot molecular cores (\citealt{mangum:1993,olmi:1996,tak:2000}), which are warm ($\geq100$\, K) and
dense ($\geq10^5-10^6$\,cm$^{-3}$) (\citealt{kurtz:2000}). Both molecules can
be used as a tracer of ongoing star formation. 

\mecn\ is a symmetric-top molecule where each rotational level is split for different projections of angular momentum along the symmetry axis of the molecule labeled by the quantum number $K$. 
Within one rotational level, $J+1 \rightarrow J$, the transitions of the $K$ components are determined solely by collisions, and their relative intensities are therefore related to the kinetic temperature of the region (\citealt{solomon:1971}). The $K$ components with one rotational level are very closely spaced in frequency and can be observed simultaneously with one backend with a bandwidth $\gtrsim$30\,MHz. 
Under the assumption that the \mecn(5--4) $K$=0, 1, 2, 3, and 4 emission is optically thin and uniformly fills the beam and that all $K$ transitions within this $J$ level are characterized by one temperature, we derived the rotational temperatures, $T_\mathrm{rot}$, and the total \mecn\ column density following \citet{araya:2005}. 
These authors studied the \mecn\ emission toward objects similar to ours (massive star-forming clumps) and showed that it is reasonable to assume the optically thin limit for this line. A beam-filling factor smaller than unity would affect the $K$ levels similarly and hence would have no large influence on the temperature derivation if the filling factor does not vary from one $K$ component to another. The \mecn\ column density, however, is affected by the beam-filling factor. Given the beam size of 27\arcsec\ against the average angular clump size found at 1.3\,mm, $\sim$20\arcsec, the beam-filling factor could be slightly smaller than unity.

\begin{figure}
\begin{center}
\includegraphics[angle=-90,width=9cm]{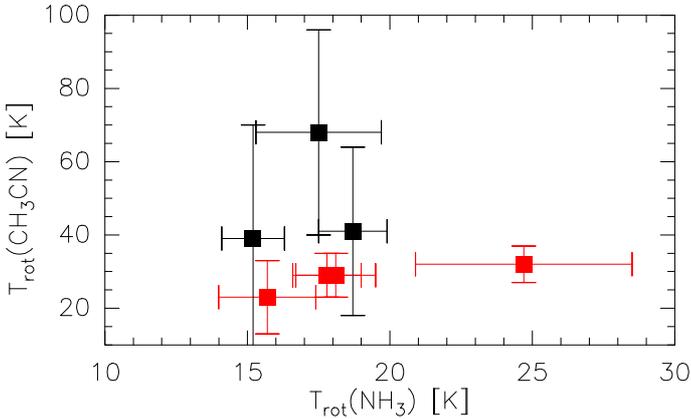}
\caption{\mecn\ rotational temperatures versus \amm\ rotational temperatures. Clumps for which the \mecn\ temperature had errors smaller than 12\,K are red. \label{fig:trotch3cn}}
\end{center}
\end{figure}

We detected the $K$=0 to 3 levels, or a subset of these, for several sources; the data were too noisy (1\,$\sigma$ rms of 0.9\,K) for a $K$=4 level detection. All detections, derived $T_\mathrm{rot}$, and column densities are given in Table \ref{ta:mecn}. The \mecn\ rotation temperatures were much higher than the \amm\ temperatures (see Table \ref{ta:mecn} and Fig.~\ref{fig:trotch3cn}), since the \mecn\ traces warmer gas than ammonia does. 
For the \h2co(4$_{03}$--3$_{04}$) line, the observed main beam brightness temperatures, LSR velocities, and line widths were determined from Gaussian fits and are listed in Table \ref{ta:sf}. 

\begin{figure}[!h]
\centering
\includegraphics[width=9cm,angle=-90]{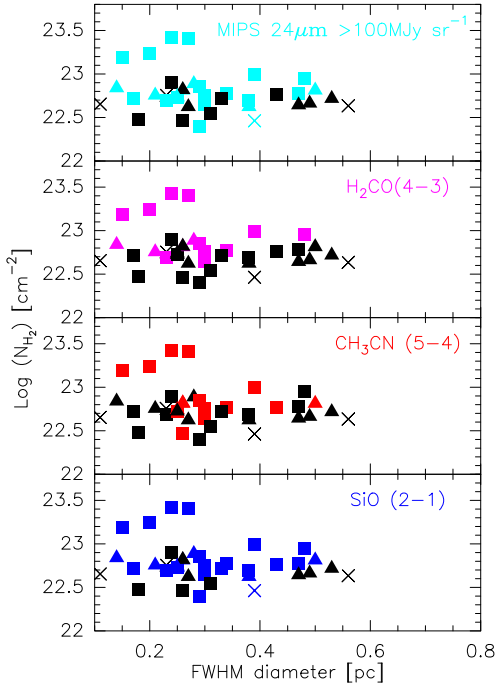}
\caption{\label{fig:sftracer} High extinction clumps placed in a diagram of column density versus the FWHM diameter (in pc) derived from the 1.2\,mm continuum (Paper I). The detections of MIPS 24\,$\mu$m emission, \h2co(4$_{03}$--3$_{04}$), \mecn(5--4) $K$=0, and SiO(2--1) are marked in color, black symbols are non-detections. The crosses mark the clumps in diffuse clouds, triangles clumps in peaked clouds, and squares clumps in multiply peaked clouds.}
\end{figure}

To understand if YSOs are preferably present in denser or more compact
clumps, we plotted all the clumps with evidence for an embedded YSO in a size versus the column density graph (Fig.~\ref{fig:sftracer}). In addition to plotting clumps with evidence of a hot molecular core through the detection of \h2co\ and \mecn , we plot clumps with a 24\,$\mu$m detection (direct evidence of dust heated by an YSO) and SiO emission (evidence of outflows, see Sect.~\ref{sect:sio}). Apparently, YSOs are found in clumps of all sizes, but, more interestingly, they are mostly found at high column densities. The minimal $N_\mathrm{H_2}$ at which the \h2co(4${03}$--3$_{04}$) line was detected is $4\times10^{22}\,\mathrm{cm^{-2}}$, while the \mecn(5--4) $K=0$emission was still observed toward one less dense clump with $N_\mathrm{H_2}=3\times10^{22}\,\mathrm{cm^{-2}}$. The column density cut-off is most clearly seen in the 24\,$\mu$m and SiO line detections; above a column density of $5\times10^{22}\,\mathrm{cm^{-2}}$, almost all clumps show infrared emission and signs of outflows.

For both the \mecn\ and \h2co\ molecules, there is a general cut-off in the column density at $4\times10^{22}\,\mathrm{cm^{-2}}$, below which no YSOs are found. This cut-off is slightly lower than the column density of $\sim$$1\times10^{23}\,\mathrm{cm^{-2}}$  theoretically required to form massive stars (0.7 g\,cm$^{-2}$, \citealt{krumholz:2008}). 
 \citet{sepulcre:2010} found a column density threshold of $7.7\times10^{22}\,\mathrm{cm^{-2}}$, below which the outflow rate decreases rapidly and the outflows are less massive. Hence, while observational data seem to strongly support a column density cutoff for (massive) star formation, they also seem to indicate a slightly lower threshold than theoretically advocated. The difference, however, is not too large, and the observed column densities are beamaverages, which means that if the source size is smaller, we are underestimating the true column density. 
 
\subsubsection{Outflow: SiO(2--1) emission}
\label{sect:sio}
Apart from evidently heating their environments, (massive) YSOs drive strong molecular outflows
(\citealt{beuther:2002b}), which may remove, in the case of disk accretion, the angular momentum away from the forming
star. While mechanisms to accelerate the outflow have been proposed for low-mass star formation (\citealt{shu:2000}), in the high-mass scenario the acceleration mechanism is not well understood yet.

Interstellar dust grains contain silicates (\citealt{draine:2003}). 
Shock waves driven by outflows can sputter Si-bearing compounds in dust grains leading to increased SiO abundances in the post-shock gas (\citealt{schilke:1997}). 
Thus, SiO is expected to be associated with ongoing star formation via its formation history. Observations support this, since
SiO is more common in more active (or evolved) sources, while very weak in
quiescent (hence early) sources (\citealt{beuther:2007b}).
Naturally, one also expects to observe wide line widths. We found line widths
from $\sim$1.5--7\,km~s$^{-1}$ for very weak SiO detections, while for stronger SiO emitting sources the line widths increase up to 26\,km~s$^{-1}$. The individual values for all detections are given in Table \ref{ta:sf} along with the main beam brightness temperatures and LSR velocities.
The mean noise level of the SiO observations is 0.07\,K,
which is comparable to previous studies in massive star-forming
regions by \citet{beuther:2007b} and \citet{motte:2007}, which reached noise levels of
0.02\,K and 0.10\,K, respectively. 

\begin{figure}
\centering
\includegraphics[angle=-90,width=7cm]{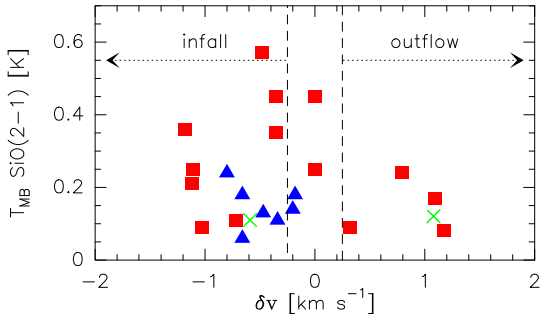}
\caption{\label{fig:sio-infall} SiO detections against the \hcop(1--0) skewness
  parameter $\delta v$. The green crosses represent sources in diffuse clouds, the blue triangles clumps in peaked clouds, and the red squares sources in multiply peaked clouds.}
\end{figure}
 
We detected the SiO(2--1) transition toward 50\% of the whole sample. 
Of all the SiO-emitting sources, a remarkable majority, 70\%, coincided with infall sources (having a \hcop(1--0) skewness parameter, $\delta v\leq -0.25$), as illustrated in Fig.~\ref{fig:sio-infall}. This finding indicates that accretion is accompanied by collimated outflows, which cause the enhancement of SiO.
A further 15\% of the SiO detections are sources exhibiting expansion ($\delta v\geq 0.25$), and the remaining 15\% are sources showing neither
expansion nor infall ($-0.25 < \delta v < 0.25$). Clumps without SiO detections have \hcop(1--0) skewness parameters divided almost equally between infall (ten clumps) and expansion (eight clumps).

\begin{figure}
\centering
\includegraphics[angle=-90,width=7cm]{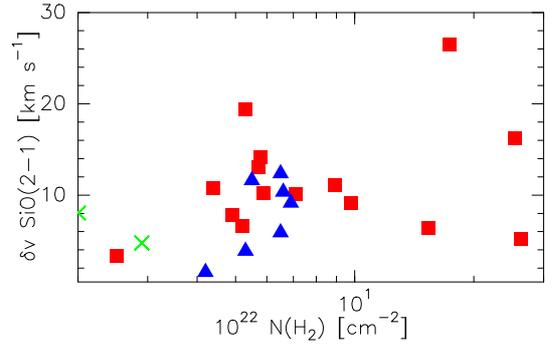}
\caption{\label{fig:sio} SiO line width against the hydrogen
  column density $N_\mathrm{H_2}$. The green crosses represent sources in diffuse clouds, the blue triangles clumps in peaked clouds, and the red squares sources in multiply peaked clouds}
\end{figure}
To check if the SiO line widths, and hence the outflows, are stronger for the more evolved sources, we compared them with the hydrogen column density determined from the 1.2\,mm continuum. Figure~\ref{fig:sio} shows that up to column densities of $10^{23}\,\mathrm{cm^{-2}}$ there is a tendency of increasing line widths with column density. The few clumps that are above this column density limit, G014.63--00.57 MM1, G017.19+00.81 MM2, G022.06+00.21 MM1, and
G014.63--00.57 MM2, show no clear trend. 
The SiO emission was detected toward all multiply peaked clouds, but just toward three peaked clouds (25\%) and only two diffuse clouds (18\%), singling out the multiply peaked clouds as active regions of star formation.

\subsubsection{Prestellar and starless sources}

Our sample contained 19 sources with no evidence of a YSO. Of these
sources, one-third shows infall in the \hcop(1--0) line profile indicating that
these are in a {\em prestellar phase}, one-third is expanding according to the \hcop(1--0) skewness parameter, and the last third shows neither infall nor outflow.
The sources without any signs of infall or YSOs were defined as {\em starless}.

\citet{motte:2007} showed how difficult it is to
find high-mass analogies of low-mass starless cores, such as L1521F
(\citealt{shinnaga:2004}), in their study of the Cygnus X complex. Remarkably, they found
SiO emission in all of the observed infrared-quiet massive dense cores,
indicating that star formation is already ongoing. 

In our sample, we found six prestellar and six starless sources. Only half of them
contain a clump (an overdensity of typical hydrogen column density of $10^{22}\,\mathrm{cm^{-2}}$, which could be identified by the Miriad routine ``sfind", see Paper I); most of them are too diffuse in the mm continuum. The masses of the prestellar clumps, derived from the 1.2\,mm continuum, were greater than their virial masses, indicating 
that gravity is the dominant force. However, also for the starless clumps, their virial parameters indicate that gravity dominates their internal dynamics. Apparently, it is not merely the lack of gravitational force that separates the starless from the prestellar clumps. The prestellar and starless sources were found in the diffuse and peaked clouds, but not in the multiply peaked clouds, which agrees with the previous indications of the multiply peaked clouds as the most evolved clouds. 
Most of the starless sources were in diffuse clouds, which are dark in the 24\,$\mu$m MIPS images. 
The prestellar sources were generally found in the vicinity of infrared objects, most likely more evolved YSOs. These YSOs are possibly responsible for introducing shocks into their surrounding medium through stellar winds and outflows, thus supporting the formation of higher densities regions  where gravitational collapse can occur. 
This implies that stars form close to
other stars, which is a well-known fact and agrees with the observed
clustering of newly formed stars (see, for example, \citealt{lada:2003}). The limited number of both starless and prestellar sources in a total of 12 clumps, compared to the 31 sources with a YSO indication, agrees with the findings of \citet{motte:2007} that the starless and prestellar phases are short and dynamic.

\subsection{CO depletion}
As a clump contracts, the level of depletion of C-bearing molecules like CO
or CS increases (\citealt{kramer:1999,bergin:1997}). The CO depletion in a clump can therefore be used
as a time marker. 
Observations of massive protostellar objects (\citealt{purcell:2009}) and
clumps in IRDCs (\citealt{pillai:2007}) show depletion of CO, indicating that
these objects are already in a state of collapse and not in the chemically earliest stage. CO depletion as a time marker should be treated with some caution, because heating of the gas (by a YSO) can cause the CO to be desorbed to the gas phase, creating low CO-depletion levels as in the chemically earliest stages around the onset of collapse.

We used \c18o, a less common isotopologue of CO, to measure the
depletion in the clumps of the high extinction clouds. In contrast to CO, the line profile of \c18o lines consisted of
one Gaussian without any wings or self-absorption. This is an indication that the \c18o(2--1)
emission has a low or moderate optical depth, which would agree with
previous observations that derived the \c18o\ optical depths by comparing the \c18o\ data with
emission from the even rarer $\mathrm{C^{17}O}$ (\citealt{crapsi:2005,pillai:2007}). Therefore, in this work, we
will assume that the \c18o(2--1) line is optically thin. The \c18o(2--1) line profiles
were fitted with a Gaussian. Table \ref{ta:c18o-ii} lists our obtained integrated intensities, the LSR velocities, and the line widths.

The column density of \c18o was calculated in the optically thin limit (see Eq.\ref{eq:c18o}) under the assumption of a beam filling factor of unity, which is likely for a
beam of 11\arcsec\ and clumps that have an average size of 20\arcsec\ in the
1.2\,mm continuum. Second, we assume LTE and equate the excitation temperature to the
rotational temperature, $T_\mathrm{rot}$, derived from the \amm(1,1) and
\amm(2,2) transitions. In very dense and cold regions, where \amm\ inversion
transitions are dominated by collisions rather than by radiative excitations and
de-excitations, $T_\mathrm{rot}$ is a good measure for the kinetic temperature
of the system (\citealt{malcolm:1983,danby:1988}). The background temperature was taken at 2.73\,K.
With the total column density of \c18o, $N_\mathrm{C^{18}O}$, known, we
calculated the abundance of \c18o\ relative to H$_2$, $\chi_\mathrm{C^{18}O}$, by comparing it to the
hydrogen column density:
\begin{equation}
\chi_\mathrm{C^{18}O}=\frac{N_\mathrm{C^{18}O}}{N_\mathrm{H_2}}\,.
\end{equation}
The canonical number for $\chi_\mathrm{C^{18}O}$ in the nearby ISM is $1.7\times10^{-7}$ (\citealt{frerking:1982}). At larger distances towards the molecular ring and the Galactic center, this value is expected to be higher (\citealt{wilson:1994}), which would increase the CO depletion we measured. 
The ratio between the observed abundance, $\chi_\mathrm{obs}$, and the
canonical value for the abundance, $\chi_\mathrm{canonical}$, then resulted in $\eta$ the CO depletion factor:
\begin{equation}
\eta=\frac{\chi_\mathrm{canonical}}{\chi_\mathrm{obs}}\,.
\end{equation}

\begin{figure}
\centering
\includegraphics[angle=-90, width=6cm]{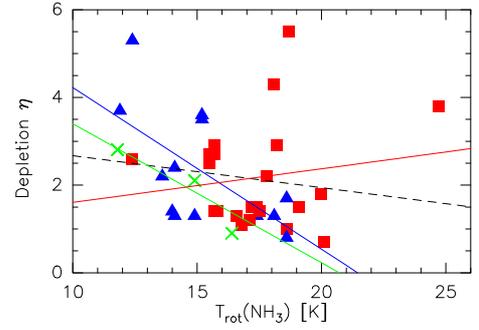}
\caption{\label{fig:eta-t} CO depletion versus the
  rotational temperature derived from \amm . The green crosses represent
  sources in diffuse clouds, the blue triangles clumps in peaked clouds, and
  the red squares sources in multiply peaked clouds. The black dashed line
  marks the trend of an increase in CO depletion with colder temperatures for all the clumps, while the green, blue, and red solid lines mark the trends for the sources in diffuse, peaked, and multiply peaked clouds.}
\end{figure}

All derived quantities, $N_\mathrm{C^{18}O}$, $N_\mathrm{H_2}$,
$\chi_\mathrm{C^{18}O}$, and $\eta$ are listed in Table \ref{ta:c18o}. In most of the clumps, \c18o\
appears to be depleted. We find a CO depletion of $\eta$=0.5--5.5 with an average of 2.3,
which is slightly lower than what is found in clumps of IRDCs ($\eta$=2--10, \citealt{pillai:2007}). 
Observations by, e.g., \citet{kramer:1999} and \citet{bacmann:2003} have shown that molecular depletion is enhanced at lower temperatures. In Fig.~\ref{fig:eta-t}, we plot for all clumps the CO depletion against the rotational temperature. While \citet{kramer:1999} found an strong exponential dependence of CO depletion with temperature, our data are too scattered to differentiate between linear and exponential behavior. Nevertheless we can see that higher $\eta$ corresponds on average to colder sources (dashed black line).

Figure \ref{fig:eta-t} shows that clumps in diffuse clouds (green crosses and green line) and peaked clouds (blue triangles and blue line) indeed follow
the trend of higher CO depletion with colder temperatures. For clumps in multiply peaked clouds (red squares and red line), the situation is different; the clumps have a large spread in temperature as well as $\eta$, and on average the opposite trend is seen: increasing $\eta$ with higher temperature. This is mainly caused by a few clumps with high temperature and high $\eta$, G017.19+0.81 MM2, G014.63--00.63 MM1, and G022.06+0.21 MM1, all
three of which have dust column densities above $10^{23}\,\mathrm{cm^{-2}}$ and water
masers. Possibly these sources were heated so fast by the embedded YSO that
the CO was not freed into the gas phase yet and the depletion values are thus still high.

\section{Discussion}

\subsection{Infall velocities}

In Sect.~\ref{sect:infall}, we determined by binomial tests that the \hcop(1--0)
line was
the best suited of the molecular transitions that we had observed to study infall. For sources that had a double-peaked \hcop(1--0) line profile, indicating a systemic infalling motion, we calculated the infall velocity $v_\mathrm{in}$ according to
\citet{myers:1996} as
\begin{equation}
v_\mathrm{in}= \frac{\Delta v_\mathrm{thin}^2}{v_\mathrm{red} - v_\mathrm{blue}}\ln\Big(\frac{1+e^{(T_\mathrm{blue}-T_\mathrm{dip})/T_\mathrm{dip}}}{1+e^{(T_\mathrm{red}-T_\mathrm{dip})/T_\mathrm{dip}}}\Big),\,
\label{eq:vin}
\end{equation}
where $v_\mathrm{red}$ and $v_\mathrm{blue}$ are the LSR velocities of the red and
blue peaks, whose main beam temperatures are given by $T_\mathrm{red}$ and
$T_\mathrm{blue}$, respectively, and $T_\mathrm{dip}$ is the
main beam temperature of the dip between the two peaks. All these quantities refer to the spectral shape of the optically thick line, except for $\Delta v_\mathrm{thin}$.
With the volume density of the clump $n_\mathrm{H_2}$ and clump radius $R$, the infall velocity can be converted to a mass infall rate $\dot{M}_\mathrm{in}$ (after \citealt{myers:1996}):
\begin{equation}
\label{eq:min}
\dot{M}_\mathrm{in}=4\pi R^2 n_\mathrm{H_2} \mu m_\mathrm{H}v_\mathrm{in}\,, 
\end{equation}
where $\mu=2.33$ the mean molecular weight and $m_\mathrm{H}$ the mass of an hydrogen atom. For the clump $n_\mathrm{H_2}$ and clump radius $R$, we took the volume density and clump radius determined in Paper I from 1.2\,mm dust continuum observations.

For low-mass star formation, the mass infall rates onto a protostar are between
 a few times $\sim10^{-5}\,M_\odot~\mathrm{yr^{-1}}$ (\citealt{palla:1993,kirk:2005}) and
$\sim10^{-6}\,M_\odot~\mathrm{yr^{-1}}$ (\citealt{whitney:1997}). This is too low
to form high-mass stars, since their prestellar phase is much shorter, less
than $3\times10^4\,$yr, according to \citet{motte:2007}, which would mean a mass infall rate of $3\times10^{-4}\,M_\odot~\mathrm{yr^{-1}}$ to form a 10$,\,M_\odot$ star. 
Theoretical models
predict a mass accretion onto a high-mass protostellar core (at radii $<0.1\,$pc) of $\dot{M}_\mathrm{in}$ of $10^{-3}\,M_\odot~\mathrm{yr^{-1}}$
(\citealt{mckee:2003,banerjee:2007}). Observations of massive clumps (with typical radii $>0.1$\,pc) find
mass infall rates of $10^{-3}-10^{-4}\,M_\odot~\mathrm{yr^{-1}}$
(\citealt{fuller:2005,peretto:2006}). In high-resolution observations of high-mass protostellar cores, the mass infall rate is indeed higher and covers the predicted order of magnitude 
($10^{-2}-10^{-4}\,M_\odot~\mathrm{yr^{-1}}$, \citealt{beltran:2006a,beltran:2008}). Our mass infall results, given in Table
\ref{ta:infallv}, fall in the range of the observed values for massive star-forming clumps. 

To explore further the mass infall rate, we calculated the infall timescale using the clump mass obtained from the 1.2\,mm
continuum (Paper I): $\tau_\mathrm{in}=\frac{M}{\dot{M}}$\,yr. This timescale can be compared with the free-fall timescale 
$\tau_\mathrm{ff} = 3.3\times10^7 \mathrm{n}^{-1/2}$\,yr, which was calculated using the density obtained from the same 1.2\,mm continuum observations (Paper I). Both timescales are given in Table~\ref{ta:infallv}. In general, the free-fall timescale is longer than the estimated infall timescale, which indicates that the mass infall rate calculated in Eq.~\ref{eq:min} is likely overestimated. The formula in Eq.~\ref{eq:min} assumes a spherically symmetric infall of the entire clump, while most likely the \hcop(1--0) emission is originating in an optically thick low-density layer in the outer parts of the clump (see also \citet{sepulcre:2010}). We also calculated the \hcop(4-3) infall velocities and mass infall rates for the few sources that showed a double-peaked blue-excess line profile (Table~\ref{ta:infallv}). Unfortunately, the error in the infall velocity of G024.94--00.15\,MM1 was quite high, leaving only G014.63--00.57\,MM1 to serve as a comparison. For the latter, the infall velocity derived from the higher excitation \hcop(4--3) line profile is half of the one derived from the \hcop(1--0) line profile, and the \hcop(4--3) infall timescale is twice the \hcop(1--0) timescale. At least for this one source, we find an indication that the \hcop(1--0) infall velocities and mass infall rates are indeed overestimated.

\begin{table*}
\centering
\caption{Infall velocities, mass infall rates, and timescales for the blue excess sources \label{ta:infallv}}
\begin{tabular}{llccccccr}
\hline\hline
\noalign{\smallskip}
Source & &\multicolumn{1}{c}{$v_{\mathrm{in10}}$}& \multicolumn{1}{c}{$\dot{M}_{\mathrm{in10}}$} & \multicolumn{1}{c}{$\tau_\mathrm{in10}$} & \multicolumn{1}{c}{$v_{\mathrm{in43}}$}& \multicolumn{1}{c}{$\dot{M}_{\mathrm{in43}}$} & \multicolumn{1}{c}{$\tau_\mathrm{in43}$} &$\tau_\mathrm{ff}$\\
       & &\multicolumn{1}{c}{($\mathrm{km~s^{-1}}$)} & \multicolumn{1}{c}{($10^{-3} M_\odot~\mathrm{yr^{-1}}$)} & \multicolumn{1}{c}{($10^5$\,yr)} & \multicolumn{1}{c}{($\mathrm{km~s^{-1}}$)} & \multicolumn{1}{c}{($10^{-3} M_\odot~\mathrm{yr^{-1}}$)} & \multicolumn{1}{c}{($10^5$\,yr)} &($10^5$\,yr)\\ 
\noalign{\smallskip}
\hline
\multicolumn{6}{l}{{\em Diffuse clouds}}\\
G013.28--00.34... &MM1& 1.09(0.15) & 1.5 & 0.6 & - & - & -&1.5 \\
G013.97--00.44... &MM1& 0.31(0.13) & - & - & -& - & - & -\\
\multicolumn{6}{l}{{\em Peaked clouds}}\\
G013.91--00.51... &MM1& 0.69(0.10) & 2.2& 0.66 & - & - & -&0.70\\
G024.94--00.15... &MM1& 1.34(0.12) &3.3 &0.32 &0.34(0.66) & 0.85 & 1.2&0.74 \\             
                                 &MM2& 3.00(0.13) &6.2 &0.13& - & - & -& 0.78\\
G030.90+00.00... &MM2& 1.76(0.15) &9.4&0.75& - & - & -&1.50\\
G035.49--00.30... &MM1& 7.4(0.13) &16.6 &0.06 & - & - & -& 0.78\\
                                &MM2& 0.84(0.10) &1.4 &0.5& - & - & -&0.85\\
G053.81--00.00... &MM1& 7.4(0.10) &9.6 &0.03& - & - & -&0.55 \\
\multicolumn{6}{l}{{\em Multiply peaked clouds}}\\
G014.63-00.57... & MM1 & 2.7(0.14) & 22.0 & 0.15& 1.30(0.35) & 10.6 & 0.30&  0.38 \\ 
                                & MM2 & - & - & - &0.90(0.28) & 2.7 & 0.26 & 0.38\\
G017.19+00.81... &MM2& 0.66(0.20) &3.1 & 0.46& - & - & -&0.42\\
                                &MM3& 1.65(0.13) &2.8& 0.26& - & - & -&0.85\\
G018.26--00.24... &MM3& 6.1(0.18) &16.0& 0.08& - & - & -&0.80\\

\noalign{\smallskip}
\hline
\end{tabular}
\tablefoot{Columns are (from left to right) source name, infall velocity; mass accretion rate; accretion timescale based on the \hcop(1--0) line, infall velocity; mass accretion rate; accretion timescale based on the \hcop(4--3) line, and free-fall timescale.}
\end{table*}

\subsection{Evolutionary sequence}
\label{sec:evodis}
In Paper I, we defined three classes of high extinction clouds, based on the contrast of the clump to the cloud in the 1.2\,mm continuum and the number of clumps per cloud. These clouds comprise diffuse clouds, clouds with a low clump-to-cloud column density contrast $C_{N_\mathrm{H_2}}$$<$2, peaked clouds, clouds with higher contrast 2$<$$C_{N_\mathrm{H_2}}$$<$3 having at least one clump, multiply peaked clouds, clouds with still higher contrast $C_{N_\mathrm{H_2}}$$>$3 that generally contained several clumps. 
Here, we review all the results in light of the hypothetical evolutionary sequence we proposed, i.e., from the quiescent diffuse clouds to the peaked ones, and finally to the multiply peaked clouds. 

\subsubsection{Diffuse clouds}

For the class of diffuse clouds, more than half of the investigated sources were not well-defined clumps but
positions of diffuse emission within a cloud. The skewness parameter analysis found a slight
excess of blue sources, indicating infall, but the binomial test showed that
this result is not significant (see Table \ref{ta:bin}). The diffuse clouds are likely the phase where the clumps are beginning to collapse. 
This would agree with the clump masses found in Paper I, where on average we found less massive clumps in the diffuse clouds than in the peaked or multiply peaked clouds (see also Table\,\ref{ta:sample}). Moreover, few YSO/hot core indications were found toward clumps in diffuse clouds, and half of the starless and prestellar sources were found toward the diffuse clouds (the other half was found toward the peaked clouds). 
A young evolutionary stage can also be identified by the CO depletion, since CO depletion requires a certain density ($>$ few$10^4\,\mathrm{cm^{-3}}$, \citealt{kramer:1999,bergin:2007}). Low depletion levels can indicate that the clump has not yet reached a high enough density. The mean depletion level of clumps in diffuse clouds ($\eta$=1.9) was slightly lower than that found toward the clumps in peaked and multiply peaked clouds ($\eta$=2.3). The depletion timescale, $\tau_\mathrm{d}\sim5\times10^9/n_\mathrm{{H_2}} ($\citealt{bergin:2007}), is $\sim$$4\times10^4$\,yr for clumps in diffuse clouds, which is slightly longer than that for in peaked clouds ($\sim$$3\times10^4$\,yr) and multiply peaked clouds ($\sim$$2\times10^4$\,yr).  

\begin{table}[!h]
\centering
\caption{\label{ta:bin} Infall properties from the \hcop(1--0) line profile for different classes of clouds}
\begin{tabular}{lrrrrrrr}
\hline\hline
\noalign{\smallskip}
Class & $N_\mathrm{blue}$ & $N_\mathrm{red}$ & $N_\mathrm{skew}$ & $N_\mathrm{tot}$ & $E$     &$P$  &  $<\delta v>$\\
\noalign{\smallskip}
\hline
\noalign{\smallskip}
Diffuse & 4               & 2          &6    &  8             &  0.25   & 0.34 & $-0.001$\\ 
Peaked & 11                & 2         &13      &  17              &  0.53   & 0.01 & $-0.63$\\
Multiply p. & 9           & 8      &17       &  20              & 0.05     & 0.50 & $-0.72$\\
\noalign{\smallskip}
\hline
\end{tabular}
\tablefoot{Columns are (from left to right) number of blue excess sources, number of red excess sources, total number of sources with significant \hcop(1--0) skewness, total number of sources, the excess parameter, the probability of the distribution to arise by chance, and the average \hcop(1--0) skewness.}
\end{table}

\subsubsection{Peaked clouds}

In the peaked clouds, the distribution of blue and red sources from the
\hcop(1--0) line profile significantly favors the blue sources with infall. The excess
parameter was 0.53 with only a 1\% possibility for this distribution to arise by
chance (Table \ref{ta:bin}). Apparently, most clumps in the peaked clouds are accreting material. The infall velocities and mass accretion rates for peaked (as well as the one for multiply peaked) clouds have a very wide range, whose mean lies far above the infall velocity/mass accretion rate of the diffuse clouds. 
All the SiO detections of clumps in peaked clouds coincide with sources that show infall based on the \hcop(1--0) skewness parameter (Fig.~\ref{fig:sio-infall}). The strong correlation between the infall characteristics and SiO outflows shows that accretion is accompanied by collimated outflows, which leads to an enhancement of SiO. This correlation does not exist for the earlier or later classes. 

\subsubsection{Multiply peaked clouds}

Among the multiply peaked clouds, there were almost equal numbers of blue and red excess sources (based on the \hcop(1--0) skewness). These clouds seem to be hosting clumps in a variety of evolutionary stages: from clumps with still ongoing accretion to clumps where the feedback from
the formed young stars halts further accretion.

Many clumps in these clouds presented SiO emission. Figure \ref{fig:sio-infall} shows that the number of clumps with SiO
emission showing infall is about equal to those showing expansion (both infall and expansion characteristics are based on the \hcop(1--0) skewness parameter), meaning that the outflow continues even after large-scale infall has finished.

The volume density of the multiply peaked clouds,
$n_\mathrm{H_2}>10^5\,\mathrm{cm}^{-3}$, is much larger than what is found for the peaked clouds. 
Multiply peaked clouds have the highest detection rate of the \mecn(5--4) $K$=0 and
\h2co(4$_{03}$--3$_{04}$) lines, indicating that this class contains the most YSOs or hot
molecular cores. In Paper I, we already found that the \amm\ rotational temperatures were on average higher for clumps in multiply peaked clouds than for the other classes (Table\,\ref{ta:sample}). 
The $K$ levels of the \mecn(5--4) line allowed a handful of rotational temperature estimates, ranging
between 23 and 34\,K. 
The increasing evidence for YSOs is likely the reason for the broadening of the line
width found toward clumps in this class (observed for the \n2hp(1--0),
\amm(1,1), \c18o(2--1), and \hh13cop(1--0) lines) by increasing turbulence.  

Finally, three clumps in the multiply peaked clouds, G014.63-00.57 MM1, G017.19+00.81 MM2, and G022.06+00.21 MM1, all show various characteristics that are different to the properties of the bulk of the sample. Examples include their SiO intensity against column density, the CO depletion against temperature, and their \n2hp(1--0)/\n2hp(3--2) ratio. While these clumps have strong outflows, suggested by their SiO emission and the presence of water masers, they also have very high column densities (theoretically enough to form massive stars, e.g., \citealt{krumholz:2008}). They show strong CO depletion values while having elevated temperatures ($\sim$ 20\,K) due to the rapid heating of the environment by the YSO. Most likely, these three clumps are actively forming massive YSOs, which is why they have such extreme properties.

\section{Summary}

We performed spectral line observations of several molecules using the IRAM 30m and the APEX telescopes toward a sample of clumps in high extinction clouds spanning various evolutionary stages. 
The skewness parameter was used to classify sources based on their line
profiles as infall sources or outflow/expanding sources. The \hcop(1--0) and
CO(3--2) lines were found to be more sensitive to infall and outflow than the
\hcop(4--3) line. Only the \hcop(1--0) profiles showed a significant excess of
infall sources, and therefore the \hcop(1--0) line profile was found to be
most effective to investigate infall profiles. 
 
SiO emission, which is thought to be a tracer of shocks, was predominantly detected toward
sources with infall. This finding is consistent with a scenario in which
accretion is accompanied by collimated outflows that lead to an enhancement
of SiO.

Between the three cloud classes, the peaked clouds have a significant high
excess of infall sources, while for the other classes no significant excess
was found. Possibly, the infall phase
has not yet begun in most diffuse clouds and already has come to an end in
some of the clumps of the multiply peaked clouds.
The infall velocities and mass infall rates suggest an increasing trend when going from the diffuse clouds (1.5 $\times10^{-3}\,M_\odot~\mathrm{yr^{-1}}$) to the peaked (and multiply peaked) clouds (1.5--16$\times10^{-3}\,M_\odot~\mathrm{yr^{-1}}$) suggesting that mass infall increases in these later stages. More measurements of clumps in diffuse clouds are, however, necessary to confirm this.
 
From RADEX calculations, we found that the ratio of the \n2hp(1--0) to
\n2hp(3--2) integrated intensity is sensitive to the hydrogen volume density, and together with the \n2hp(1--0) optical depth can be used to derive good estimates of not only the hydrogen volume density but also the intrinsic \n2hp\ column density. 
Apart from the ammonia rotational temperatures, determined in Paper I, we added several other temperature estimates. All estimates converge to a temperature range between 10 and 40\,K, where the higher value is determined by \mecn, a molecule usually found in hot cores and UCH{\sc ii}\,regions. Similarly, various density estimates place the density in the order of $10^5\,\mathrm{cm^{-3}}$.

The presence of YSOs in the clumps of the high extinction clouds was established by 24\,$\mu$m emission, but also indirect evidence through outflows (SiO emission) and the presence of hot molecular cores through \mecn\ and \h2co\ emission was gathered.
We find that most of the YSOs are located in clumps of peaked and multiply peaked clouds, while almost no YSO indications were present in diffuse clouds. The presence of an YSO seems independent of the physical size of the clump, since YSOs are found over a large range of clump diameters. However, there seems to be (hydrogen) column density cutoff at $(4-5)\times10^{22}\,\mathrm{cm^{-2}}$, below which few or no YSOs are found. The cutoff is most clear in the 24\,$\mu$m emission and in the presence of SiO emission. 

In summary, the diffuse clouds seem to contain clumps that are on the verge of infall. The first stages of star formation are ongoing in the peaked clouds, since a majority of the clumps is infalling and many clumps also show outflows. The multiply peaked clouds harbor the more evolved clumps, where infall has already stopped in some of them, and generally we find indications of YSOs. Hence, the cloud morphology seems to have a direct connection with the evolutionary stage of the objects forming inside. 

\begin{acknowledgements}
We would like to express our thanks to Floris van der Tak and Malcolm Walmsley for their constructive comments that lead to a much improved manuscript. 
We are grateful for the help and assistance of the staff of both the IRAM 30m and the APEX 12m telescopes during the observations. KLJR is funded by an ASI fellowship under contract number I/005/11/0. 
\end{acknowledgements}
\bibliographystyle{aa}
\bibliography{/Users/kazi/tot}
\appendix
\section{Formulas and equations used}
\subsection{N$_2$H$^+$}
\label{app:n2hp}
The excitation temperature $T_\mathrm{ex}$ can be calculated as
\begin{equation}
T_\mathrm{ex}=\frac{h\nu/k}{ln[\frac{h\nu/k}{A/(\tau_\mathrm{m} f)} +1]},
\end{equation}
where $A=\tau_\mathrm{m}f(J_\nu(T) +J_\nu(T_\mathrm{bg})$ is the outcome of the CLASS hfs method fitting. Using $ln(\epsilon +1)\simeq \epsilon$, one can simplify the above equation as
\begin{equation}
T_\mathrm{ex}\simeq {A}{\tau_\mathrm{m} f} +1.
\end{equation}
If we ignore the ``+1" in the above equation, we find that $T_\mathrm{ex}\propto1/f$. Since the average $T_\mathrm{ex}$ was found to be around 10\,K, ignoring the ``+1" in the above equation would lead to an uncertainty of (on average) 10\%. 

We calculated the column density of \n2hp, $N_\mathrm{N_2H^+}$,  following \citet{benson:1998}:
\begin{equation}
N_{\mathrm{N_2H^+}} = 3.3\times10^{11}\frac{\tau_\mathrm{tot} \Delta v T_\mathrm{ex}}{1-e^{-4.47/T_\mathrm{ex}}}\hspace{0.5cm}\mathrm{[cm^{-2}]}\,,
\label{eq:n2hp}
\end{equation}
where $\Delta v$ is the line width obtained from the hyperfine fit and $T_\mathrm{ex}$ is the excitation temperature.

\subsection{CO}
In LTE, the total column density,  $N_\mathrm{tot}$, of the CO molecule can be calculated from the integrated main beam brightness temperature of an optically thin line as follows:
\begin{equation}
N_\mathrm{tot}=\frac{3h}{8\pi^3S\mu_D^2}\frac{Q(T_\mathrm{ex})}{g_ug_Kg_{nuc}}\frac{e^{E_u/kT_\mathrm{ex}}}{e^{h\nu/kT_\mathrm{ex}}-1}\frac{\int T_\mathrm{MB}\Delta v}{J_\nu(T_\mathrm{ex})-J_\nu(2.73)}\,,
\label{eq:c18o}
\end{equation}
\begin{tabular}{lll}
where& $h$ &Planck constant\\
     & $k$ & the Boltzmann constant\\
     & $S$ &line strength\\
     & $\mu_D$ & the dipole moment of the molecule\\
     & $T_\mathrm{ex}$ & the excitation temperature\\
     & $Q(T_\mathrm{ex})$ &partition function at $T_\mathrm{ex}$, \\
     & & $Q(T_\mathrm{ex})\simeq0.38(T_\mathrm{ex}+0.88)$\\
     & $g_u$ & the rotational degeneracy for the upper energy\\
     & $g_K$ & the K degeneracy\\
     & $g_{nuc}$ & the nuclear degeneracy\\
     & $E_u$ & the energy of the upper level\\
     & $\nu$ & the transition frequency\\
     & $J_\nu(T)$ & $\frac{h\nu/k}{e^{h\nu/kT}-1}$\,.\\
&&\\
\end{tabular}

The properties of the \c18o\ J=2$\rightarrow$1 transition are: $S$= $J_u/(2J_u +1)$=$0.4$,
where $J_u$ is the angular momentum quantum number of the upper $J$ level, $\mu_D$=$0.1098\,\mathrm{Debye}$=$0.1098\times10^{-18}$\,e.s.u., $g_u$=$2J_u +1$=$5$, $g_K$=$1$,  $g_{nuc}$=$1$,
$E_u$=$k\times15.81$\,erg, and $\nu$=$219.56$\,GHz. 

The general formula for the CO column density, valid in the optically thick and thin regime is
\begin{equation}
N_\mathrm{tot}=\frac{3h}{8\pi^3S\mu_D^2}\frac{Q(T_\mathrm{ex})}{g_ug_Kg_{nuc}}\frac{e^{E_u/kT_\mathrm{ex}}}{e^{h\nu/kT_\mathrm{ex}}-1}\int\tau\Delta v\,,
\label{eq:c18ogen}
\end{equation}
where all the quantities are as defined above. 
\section{Online Material}

\onltab{1}{
\onecolumn
\begin{landscape}
\setlength{\LTcapwidth}{1.4\textheight}
\begin{longtable}{llrrcccccccccccc}
\caption{\label{ta:det}Molecular line survey overview: `$\surd$' for a detection and `..' when not observed. For non-detections 3$\sigma$ upper limits are given in Kelvin. Positions are expressed in J2000 coordinates. Water maser detections from Paper I.}\\
\hline\hline
Source name && \multicolumn{1}{c}{R.A.}     & \multicolumn{1}{c}{Declination}  & \hcop&\hh13cop&\hcop& \n2hp& \n2hp&SiO& \mecn & H$_2$CO& CO& \c18o& CO&H$_2$O\\
     && \multicolumn{1}{c}{(h:m:s)}  & \multicolumn{1}{c}{(\degr:\,\arcmin:\,\arcsec)}  &(1--0) & (1--0) & (4--3) & (1--0) & (3--2) & (2--1) & (5--4),$K$=0 & (4$_{03}$--3$_{04}$) & (2--1)&      (2--1) & (3--2) & 22\,GHz\\
\noalign{\smallskip}
\hline
\noalign{\smallskip} 
\endfirsthead
\caption{continued.}\\
\hline\hline
Source name && \multicolumn{1}{c}{R.A.}     & \multicolumn{1}{c}{Declination}  & \hcop&\hh13cop&\hcop& \n2hp& \n2hp&SiO& \mecn & H$_2$CO& CO& \c18o& CO&H$_2$O\\
      && \multicolumn{1}{c}{(h:m:s)}  & \multicolumn{1}{c}{(\degr:\,\arcmin:\,\arcsec)} &(1--0) & (1--0) & (4--3) & (1--0) & (3--2) & (2--1) & (5--4),$K$=0 & (4$_{03}$--3$_{04}$) & (2--1)&      (2--1) & (3--2)&22\,GHz\\
\noalign{\smallskip}
\hline
\noalign{\smallskip}
\endhead
G012.73--00.58... &MM1&18:15:41.3   & --18:12:44   &$\surd$ &$\surd$ & ..           &$\surd$ & ..             & $<$0.21              & $<$0.48    & ..   &$\surd$ & $<$0.78 &.. & ..\\
                                 &MM2& 18:15:32.7   & --18:10:15  &$\surd$ &$\surd$ & ..          &$\surd$ & ..              & $<$0.26             & $<$0.54     & ..   &$\surd$ &$\surd$ &..&..\\
G013.28--00.34... &MM1& 18:15:39.9   & --17:34:37  & $\surd$ &$\surd$ & ..          &$\surd$ & ..             &$\surd$   & $<$0.47    & ..   &$\surd$ &$\surd$ &..&n.d.\\
G013.91--00.51... &MM1& 18:17:34.8   & --17:06:52  &$\surd$ &$\surd$ &$\surd$ &$\surd$ &$\surd$  & $<$0.32            & $<$0.48     &$\surd$&$\surd$ &$\surd$ &$\surd$&n.d.\\
G013.97--00.45... &MM1& 18:17:16.5   & --17:01:16  &$\surd$ &$\surd$ & ..           &$\surd$  & ..            & $<$0.25            & $<$0.48    & ..    &$\surd$ &$\surd$ &..&..\\
G014.39--00.75A.. &MM1& 18:19:19.0   & --16:43:49  &$\surd$ &$\surd$ & ..         &$\surd$  &$\surd$  & $<$0.27            & $<$0.60    &$\surd$&$\surd$ & $\surd$ &..&n.d.\\
                                  &MM2&18:19:17.4   & --16:44:04  &$\surd$ &$\surd$ & ..           &$\surd$  & ..            & $<$0.20          &  $<$0.32   & ..    &$\surd$ &$\surd$ &..&..\\
G014.39--00.75B.. &MM3& 18:19:33.3   & --16:45:01   &$\surd$ &$\surd$ & ..         &$\surd$ & ..             &$<$0.21           & $<$0.34    & ..    &$\surd$ &$\surd$ &..&..\\
G014.63--00.57... &MM1& 18:19:15.2   & --16:29:59   &$\surd$ &$\surd$ &$\surd$ &$\surd$ &$\surd$ & $\surd$ &$\surd$&$\surd$&$\surd$ &$\surd$ &$\surd$&$\surd$\\
                                 &MM2& 18:19:14.3   & --16:30:41   &$\surd$ &$\surd$ &$\surd$ &$\surd$ &$\surd$ & $\surd$ &$\surd$ &$\surd$&$\surd$ &$\surd$ &$\surd$&..\\
                                 &MM3& 18:19:02.9   & --16:30:29   &$\surd$ &$\surd$ & ..           &$\surd$ & ..             &$<$0.18            & $<$0.26    & ..    &$\surd$ &$\surd$ &..&..\\
                                 &MM4& 18:19:20.5   & --16:31:42   & $\surd$ &$\surd$ & ..           &$\surd$ & ..            & $<$0.18            & $<$0.29    & ..    &$\surd$ &$\surd$ &..&..\\
G016.93+00.24... &MM1&18:20:50.8  & --14:06:01  &$\surd$ &$\surd$ & ..               &$\surd$ & ..            & $<$0.20             & $<$0.24   & ..    &$\surd$ &$\surd$ &..&..\\
G017.19+00.81... &MM1& 18:19:08.9   & --13:36:29  & $\surd$ &$\surd$ &$\surd$ &$\surd$ &$\surd$ & $\surd$  & $<$0.24    &$\surd$&$\surd$ &$\surd$ &$\surd$&..\\
                                &MM2& 18:19:12.9   & --13:33:46 & $\surd$ &$\surd$ &$\surd$  &$\surd$ &$\surd$ & $\surd$   &$\surd$&$\surd$&$\surd$ &$\surd$ &$\surd$&$\surd$\\\
                                &MM3&18:19:12.1   & --13:33:32  & $\surd$ &$\surd$ & ..             &$\surd$ & ..            & $\surd$  & $\surd$  & ..    &$\surd$ &$\surd$ &..&..\\
                                &MM4& 18:19:15.2   & --13:39:29 &$\surd$ &$\surd$ &$\surd$   &$\surd$ &$\surd$  & $<$0.14              & $\surd$     & $<$0.54  &$\surd$ &$\surd$ &$\surd$&..\\
G018.26--00.24... &MM1& 18:25:11.8   & --13:08:04  &$\surd$ &$\surd$ &$\surd$ &$\surd$ &$\surd$ & $\surd$  &$\surd$&$\surd$&$\surd$ &$\surd$ &$\surd$&n.d.\\
                                 &MM2& 18:25:06.4   & --13:08:51  &$\surd$ &$\surd$ &$\surd$ &$\surd$ &$\surd$ & $\surd$  & $<$0.30     &$\surd$&$\surd$ &$\surd$ &$\surd$&n.d.\\
                                &MM3&18:25:05.6   & --13:08:20  &$\surd$ &$\surd$ &$\surd$ &$\surd$ &$\surd$ & $\surd$    &$\surd$&$\surd$&$\surd$ &$\surd$ &$\surd$&..\\
                                &MM4& 18:25:04.5   & --13:08:27  & $\surd$ &$\surd$ &$\surd$ &$\surd$ &$\surd$ & $\surd$  & $\surd$   &$\surd$&$\surd$ &$\surd$ &$\surd$&..\\
                                &MM5& 18:25:01.8   & --13:09:06  &$\surd$ &$\surd$ & ..             &$\surd$ & ..            & $<$0.20             & $<$0.26             & ..     &$\surd$ &$\surd$ &..&..\\
G022.06+00.21... &MM1& 18:30:34.7   &  --9:34:46  &$\surd$ &$\surd$ &$\surd$  &$\surd$  &$\surd$  & $\surd$ &$\surd$   &$\surd$&$\surd$ &$\surd$ &$\surd$&$\surd$\\
                                &MM2& 18:30:38.5   &  --9:34:29  &$\surd$ &$\surd$ & ..              &$\surd$ & ..             & $<$0.21             &   $<$0.27          & ..      &$\surd$ &$\surd$ &..&..\\
G023.28--00.12... &MM1& 18:34:23.5   &  --8:32:20  &$\surd$ &$\surd$ &$\surd$  &$\surd$ &$\surd$  & $\surd$ & $\surd$   & $<$0.54       &$\surd$ &$\surd$ & $\surd$&..\\
                                 &MM2& 18:34:20.4   &  --8:33:16  & $\surd$ &$\surd$ & ..            &$\surd$ & ..            & $<$0.21              & $<$0.26                 & ..      &$\surd$ &$\surd$ &..&..\\
G024.37--00.15... &MM1& 18:36:27.8   &  --7:40:24  &$\surd$ &$\surd$ & ..             &$\surd$ & ..            & $<$0.18             & $<$0.25   & ..      &$\surd$ &$\surd$ &..&n.d.\\
                                 &MM2& 18:36:18.3   &  --7:41:00  &$\surd$ &$\surd$ &$\surd$ &$\surd$ &$\surd$ & $\surd$  & $\surd$     &$\surd$&$\surd$ &$\surd$ &$\surd$&$\surd$\\
G024.61--00.33... &MM1& 18:37:23.1   &  --7:31:39  &$\surd$ &$\surd$ &$\surd$ &$\surd$ &$\surd$ &$\surd$    & $\surd$    &$\surd$&$\surd$ &$\surd$ &$\surd$&n.d.\\
                                &MM2& 18:37:21.3   &  --7:33:07  &$\surd$ &$\surd$ & ..             &$\surd$ & ..            & $<$0.18               & $<$0.20         & ..     &$\surd$ &$\surd$ &..&n.d.\\
G024.94--00.15... &MM1& 18:37:19.7   &  --7:11:41   &$\surd$ &$\surd$ &$\surd$ &$\surd$ &$\surd$ & $\surd$   & $\surd$    & $\surd$&$\surd$ &$\surd$ &$\surd$&$\surd$\\
                                &MM2& 18:37:12.2   &  --7:11:23  &$\surd$ &$\surd$ & ..            &$\surd$  & ..             & $\surd$   & $\surd$      &  ..    &$\surd$ &$\surd$ &..&n.d.\\
G025.79+00.81... &MM1& 18:35:20.5   &  --5:56:36  &$\surd$ &$\surd$ & ..           &$\surd$    & ..            & $\surd$   &  $<$0.24            & ..     &$\surd$ &$\surd$ &..&..\\
                                &MM2& 18:35:26.3   &  --5:59:21  &$\surd$ &$\surd$ & ..          &$\surd$    & ..             & $<$0.16              & $<$0.24   &  ..    &$\surd$ &$\surd$ &..&..\\
G030.90+00.00A..&MM1&18:47:28.9   &  --1:48:07   &$\surd$ &$\surd$ &$\surd$ &$\surd$ &$\surd$  & $<$0.26              &  $<$0.27    & $<$0.56   &$\surd$ &$\surd$ &$\surd$&..\\
G030.90+00.00B.. &MM2& 18:47:41.9   &  --1:52:13  &$\surd$ &$\surd$ &$\surd$ &$\surd$ &$\surd$ & $<$0.26                 & $<$0.27    & $<$0.53      &$\surd$ &$\surd$ &$\surd$&-\\
G030.90+00.00C.. &MM3& 18:47:48.2   &  --1:51:30  &$\surd$ &$\surd$ & ..           &$\surd$ & ..              & $<$0.15               &  $<$0.20   & ..     &$\surd$ &$\surd$ &..&n.d.\\
G034.71--00.63... &MM1&  18:56:48.3   &   1:18:49 & $\surd$ &$\surd$ &$\surd$ &$\surd$ &$\surd$   & $\surd$     &$\surd$&$\surd$&$\surd$ &$\surd$ & $\surd$&n.d.\\ 
                                &MM2&18:56:58.2   &   1:18:44 &$\surd$ &$\surd$   &$\surd$&$\surd$ &$\surd$      & $\surd$     & $<$0.22          & $<$0.57 &$\surd$ &$\surd$ &$\surd$&..\\ 
                                 &MM3&18:57:06.5   &   1:16:52 &$\surd$ &$\surd$ & ..           &$\surd$   & ..                & $\surd$     &  $<$0.19            & ..     &$\surd$ &$\surd$ &..&n.d.\\ 
G034.77--00.81... &MM1& 18:57:40.7  &   1:16:09 &$\surd$ &$\surd$ & ..           &$\surd$    &  ..               & $<$0.14                 &  $<$0.20   & ..     &$\surd$ &$\surd$ &..&..\\ 
G034.85+00.43... &MM1& 18:53:23.2  &   1:53:16 &$\surd$ &$\surd$ & $<$2.8    &$\surd$          & $<$0.72                & $<$0.15                &  $<$0.20    & $<$0.63     &$\surd$ &$\surd$ &$\surd$&..\\ 
G035.49--00.30A.. &MM1& 18:57:05.2   &   2:06:29 &$\surd$ &$\surd$ & ..       &$\surd$     & ..                  & $<$0.17             & $\surd$    & ..     &$\surd$ &$\surd$ &..&$\surd$\\ 
G035.49--00.30B.. &MM2&18:57:08.4   &   2:09:01 &$\surd$ &$\surd$ & ..     &$\surd$         & ..                & $\surd$    & $<$0. 24  & ..     &$\surd$ &$\surd$ &..&n.d.\\ 
                                  &MM3&18:57:08.1   &   2:10:47 & $\surd$ &$\surd$ & ..     &$\surd$        & ..                & $\surd$     &  $<$0.22    & ..     &$\surd$ &$\surd$ &..&$\surd$\\ 
G037.44+00.14A.. &MM1& 18:59:14.0   &   4:07:37 & $<$0.12          &$\surd$ & ..     &$\surd$         & ..                 & $<$0. 18                &  $<$0.24    & ..     &$\surd$ &$\surd$ &..&..\\ 
G037.44+00.14B..&MM2& 18:59:10.2   &   4:04:32 &  $\surd$ & $\surd$ & ..     &$\surd$      & ..                & $<$0.16                 &  $<$0.24   & ..     &$\surd$ & .. &..&..\\ 
G050.06+00.06... &MM1&19:23:12.4   &  15:13:35 & $\surd$ &$\surd$ & ..     &$\surd$        & ..                & $<$0.18                 & $<$0.22   & ..     &$\surd$ &$\surd$ &..&n.d.\\
                                &MM2& 19:23:09.2   &  15:12:42  &$\surd$ &$\surd$ & ..           &$\surd$ & ..             & $<$0.20                   &  $<$0.26                 & ..     &$\surd$ &$\surd$ &..&..\\ 
G053.81--00.00... &MM1& 19:30:55.7  &  18:29:55 &$\surd$ &$\surd$ & ..           &$\surd$ &$\surd$   & $\surd$         &  $<$0.22    & $\surd$&$\surd$ &$\surd$ &.. &n.d.\\
\noalign{\smallskip}
\hline
\end{longtable}   
\end{landscape}   
\twocolumn
}

\onltab{2}{
\onecolumn
\begin{landscape}
\setlength{\LTcapwidth}{1.4\textheight}
\begin{longtable}{llccccccccc}
\caption{Infall and outflow study using the \hcop(1--0), CO(2--1), \hcop(4--3) and CO(3--2) lines. When the source was not observed the field is marked as `..'.\label{ta:infall}}\\
\hline\hline
\noalign{\smallskip}
Source name & &\multicolumn{3}{c}{\underline{\hspace{1.5cm}\hcop(1--0)\hspace{1.5cm}}} & \multicolumn{2}{c}{\underline{\hspace{1.5cm}CO(2--1)\hspace{1.5cm}}} & \multicolumn{2}{c}{\underline{\hspace{1.0cm}\hcop(4--3)\hspace{1.0cm}}} & \multicolumn{2}{c}{\underline{\hspace{1.0cm}CO(3--2)\hspace{1.0cm}}} \\
 & &$\frac{T_{\mathrm{blue}}}{T_{\mathrm{red}}}$ & $\delta v$ &Profile & Profile & Wings/Range& Profile &  $\delta v$ & Profile &$\delta v$\\
 & & &  &  &  & /(km~s$^{-1}$)& & & &\\
\noalign{\smallskip}
\hline
\noalign{\smallskip}
\endfirsthead
\caption{continued.}\\
\hline\hline
\noalign{\smallskip}
Source name & &\multicolumn{3}{c}{\underline{\hspace{1.5cm}\hcop(1--0)\hspace{1.5cm}}} & \multicolumn{2}{c}{\underline{\hspace{1.5cm}CO(2--1)\hspace{1.5cm}}} & \multicolumn{2}{c}{\underline{\hspace{1.0cm}\hcop(4--3)\hspace{1.0cm}}} & \multicolumn{2}{c}{\underline{\hspace{1.0cm}CO(3--2)\hspace{1.0cm}}} \\
 & &$\frac{T_{\mathrm{blue}}}{T_{\mathrm{red}}}$ & $\delta v$ &Profile & Profile & Wings/Range& Profile &  $\delta v$ & Profile &$\delta v$\\
 & & &  &  &  & /(km~s$^{-1}$)& & & &\\
\noalign{\smallskip}
\hline
\noalign{\smallskip}
\endhead
G012.73--00.58... &MM1&     -    & 0.19(0.01) &  -& - & y/11& ..& ..& ..& ..\\ 
              &MM2& 0.43(0.04) & 0.51(0.13)     & R &-  & y/13& ..& ..& ..& ..\\
G013.28--00.34... &MM1& 1.58(0.34) &--0.59(0.13)     & B & B & y/14& ..& ..& ..& ..\\
G013.91--00.51... &MM1& 1.63(0.17) &--0.64(0.19)     & B & R  & y/27& B & --0.46(0.21)& - & -\\
G013.97--00.45... &MM1& 1.24(0.17) &--0.73(0.13)     & B & -& n& ..& ..& ..& ..\\
G014.39--00.75A.. &MM1&  -       &--0.52(0.03) & B & B & y/18& ..& ..& ..& ..\\
             &MM2& ..         &--0.45(0.03) & B & B & y/13& ..& ..& ..& ..\\
G014.39--00.75B..               &MM3&   -      &--5.79(0.29) & B & - & y/15& ..& ..& ..& ..\\
G014.63--00.57... &MM1& 1.05(0.09) &    -     & -& B & y/30& B & --0.59(0.21)& B & --2.06(0.21)\\ 
              &MM2& 0.45(0.13) & 1.18(0.13)     & R & B& y/23& B& --0.78(0.21)&- & -\\
              &MM3&    -     & 0.55(0.04) & R & B&y/16& ..& ..& ..& ..\\
              &MM4&    -    & 0.22(0.03) & -  & -& y/12& ..& ..& ..& ..\\
G016.93+00.24... &MM1&    -      &--0.62(0.13)     & B & -&  n & ..& ..& ..& ..\\
G017.19+00.81... &MM1& 0.35(0.04) & 0.32(0.13)     & R & R & y/29& R &0.37(0.21)& R &1.07(0.21)\\  
              &MM2& 1.47(0.15) &--0.48(0.13)     & B & R &y/43& R&0.55(0.21) & R & 1.15(0.21)\\
              &MM3& 2.39(0.36) &--0.35(0.13)     & B & B & y/35& ..& ..& ..& ..\\
              &MM4& 0.40(0.08) & 0.27(0.16)     & R & R & y/14& -&--0.06(0.23)& R & 0.48(0.23)\\
G018.26--00.24... &MM1&     -     &--1.18(0.04) & B & B & y/20& R & 0.42(0.22)& B & --1.45(0.22)\\
              &MM2&     -    &--1.03(0.14)     & B & B & y/12 & -& --0.11(0.22)& B & --0.98(0.22) \\
              &MM3& 1.68(0.65) &--1.12(0.14)     & B &B  & y/25& -&0.00(0.22)& B & --1.18(0.22) \\
              &MM4& 0.89(0.35) &     -    &  -& R & y/23& -& --0.04(0.23)& B&--1.14(0.23)\\
              &MM5&     -    & 0.49(0.16)     & R & B & y/15& ..& ..& ..& ..\\ 
G022.06+00.21... &MM1& 0.23(0.07) & 0.79(0.14)     & R & R & y/46& -&--0.11(0.22)& R& 1.31(0.22)\\
              &MM2& 0.55(0.21) & 0.76(0.15)     & R & R & y/12& ..&..& ..& ..\\
G023.28--00.12... &MM1&-        &--0.47(0.14)     & B &-  & n & - &--0.23(0.22)& R&0.87(0.22)\\
              &MM2& 0.26(0.07) & 0.33(0.22)      & R & -& y/14& ..& ..& ..& ..\\
G024.37--00.15... &MM1&    &--0.59(0.22)     & B & B & y/11& ..& ..& ..& ..\\ 
              &MM2&     &--1.11(0.26) & B & R& y/17&-  & --0.15(0.31) &B&--1.11(0.31)\\ 
G024.61--00.33... &MM1&         - &--0.72(0.13)     & B & B & y/27& R& 0.26(0.22)& B &--1.75(0.21)\\
              &MM2&    -      &--0.07(0.14)     & - & B & y/40& ..& ..& ..& ..\\
G024.94--00.15... &MM1& 2.03(0.24) &--0.80(0.14)     & B & B & y/21& B& --0.43(0.28)&B & --0.99(0.22)\\
              &MM2& 2.24(0.30) &--0.66(0.16)     & B & B & y/27& ..& ..& ..& ..\\
G025.79+00.81... &MM1& 0.44(0.10) & 1.08(0.14)     & R &- & y/10& ..& ..& ..& ..\\
              &MM2&    -      & 0.25(0.06) & R &-  & n & ..& ..& ..& ..\\
G030.90+00.00A..&MM1&   -   & 0.66(0.03) & R & R & y/20& -&--0.04(0.16)& R&0.63(0.22)\\
G030.90+00.00B..&MM2& 1.90(0.24) &--0.42(0.18)     & B & B & y/13& -&--0.17(0.26)& -& --0.19(0.24)\\
G030.90+00.00C..&MM3& -   &--0.27(0.02) & B & B & y/26& .. &..& ..& ..\\
G034.71--00.63... &MM1&  -   &--0.11(0.01) & .. & .. & y/43&..& ..& ..& ..\\
              &MM2&  -    &--0.35(0.07) & B & R & y/25& B & --0.40(0.22) & R& 0.35(0.22)\\
              &MM3&  -   & 1.09(0.04) & R & R & y/22& ..& ...& ..& ..\\
G034.77--00.81... &MM1&  -  &--0.76(0.07) & B & B & n & ..& ..& ..& ..\\
G034.85+00.43... &MM1&   -  & 0.31(0.03) & R& B & n&  ..& ..& ..& ..\\
G035.49--00.30A.. &MM1& 2.78(0.49) &--0.66(0.05) & B & B & y/16& ..& ..& ..& ..\\
G035.49--00.30B..              &MM2& 1.61(0.16) &--0.20(0.01) & -& B & y/11& ..& ..& ..& ..\\
              &MM3&   -   &--0.18(0.01) & - & B& y/11& ..& ..& ..& ..\\
G037.44+00.14A.. &MM1& -      & -          & - & - & n &..& ..& ..& ..\\ 
G037.44+00.14B..               &MM2&- & 0.00(0.20)& -& -& n& ..& ..& ..& ..\\
G050.06+00.06... &MM1& 0.97(0.12) &0.0(0.14)       & -& -&y/14 & ..& ..& ..& ..\\
              &MM2&     -    &0.00(0.01) & - &- &y/17& ..& ..& ..& ..\\
G053.81--00.00... &MM1& 3.02(0.33) &--0.34(0.13)       & B & B&y/34& ..& ..& ..& ..\\
\noalign{\smallskip}
\hline
\end{longtable}
\tablefoot{Columns are (from left to right) source name, ratio of the blue over red peak intensity of the \hcop(1--0) line, \hcop(1--0) skewness parameter, \hcop(1--0) profile, CO(2--1) profile, CO(2--1) wings/CO range of wings, \hcop(4--3) profile, \hcop(4--3) skewness parameter, CO(3--2) profile, and CO(3--2) skewness parameter. }
\end{landscape}
\twocolumn
}

\onltab{3}{
\onecolumn
\begin{landscape}
\setlength{\LTcapwidth}{1.4\textheight}
\begin{longtable}{llccccccccc}
\caption{Molecular line survey results for \hcop(1--0) and \hh13cop(1--0); columns 7--11 are based on Gaussian fits.\label{ta:hcop}}\\
\hline\hline
Source name &  & \multicolumn{6}{c}{\underline{\hspace{5cm}\hcop (1--0)\hspace{5cm}}}&\multicolumn{3}{c}{\underline{\hspace{2cm}\hh13cop(1--0)\hspace{2cm}}}\\
       &  & $T_\mathrm{blue}$ &$ v_\mathrm{LSR, blue}$  & $T_\mathrm{red}$ &$ v_\mathrm{LSR, red}$ & $T$ & $v_\mathrm{LSR}$  & $T$ & $v_\mathrm{LSR}$  & $\Delta V$ \\
      &  & (K) & ($\mathrm{km~s^{-1}}$) & (K) & ($\mathrm{km~s^{-1}}$) & (K) & ($\mathrm{km~s^{-1}}$)&(K) & ($\mathrm{km~s^{-1}}$)&($\mathrm{km~s^{-1}}$\\
\noalign{\smallskip}
\hline
\noalign{\smallskip}
\endfirsthead
\caption{continued.}\\
\hline\hline
Source name &  & \multicolumn{6}{c}{\underline{\hspace{5cm}\hcop (1--0)\hspace{5cm}}}&\multicolumn{3}{c}{\underline{\hspace{2cm}\hh13cop(1--0)\hspace{2cm}}}\\
       &  & $T_\mathrm{blue}$ &$ v_\mathrm{LSR, blue}$  & $T_\mathrm{red}$ &$ v_\mathrm{LSR, red}$ & $T_{MB}$ & $v_\mathrm{LSR}$  & $T_{MB}$ & $v_\mathrm{LSR}$  & $\Delta V$ \\
      &  & (K) & ($\mathrm{km~s^{-1}}$) & (K) & ($\mathrm{km~s^{-1}}$) & (K) & ($\mathrm{km~s^{-1}}$)&(K) & ($\mathrm{km~s^{-1}}$)&($\mathrm{km~s^{-1}}$\\
\noalign{\smallskip}
\hline
\noalign{\smallskip}
\endhead
G012.73--00.58... &MM1&-&-&-&-& 1.72(0.09) & 6.82(0.01)& 0.75(0.08) & 6.69(0.02) & 0.68(0.05) \\
              &MM2&1.04(0.06) & 5.16(0.25) & 2.41(0.06) & 6.87(0.25) &-&-& 1.19(0.11) & 6.50(0.02)  & 0.72(0.05) \\
G013.28--00.34... &MM1&1.20(0.10) &40.17(0.25) & 0.76(0.10) & 42.49(0.25)&-&-& 0.55(0.06) & 41.30(0.06) & 1.91(0.15) \\
G013.91--00.51... &MM1&2.08(0.08) &22.08(0.25) & 1.27(0.08) & 24.05(0.25)&-&-& 1.7(0.07) & 22.97(0.02) & 1.39(0.04) \\
G013.97--00.45... &MM1&1.43(0.09) &18.51(0.25) & 1.15(0.09) & 20.88(0.25)&-&-& 0.82(0.06) & 20.03(0.05) & 2.08(0.11)\\
G014.39--00.75A.. &MM1&-&-&-&-& 2.39(0.12) & 17.20(0.03) & 1.02(0.08) & 17.83(0.03) & 1.22(0.07)\\
                &MM2&-&-&-&-& 0.86(0.12) & 16.99(0.06)& 0.79(0.06) & 17.58(0.03) & 1.3(0.08) \\
G014.63--00.57... &MM1&3.18(0.13) & 16.19(0.25) & 3.04(0.13) &20.30(0.25) &-&-& 1.25(0.05) & 18.55(0.03) & 2.45(0.07)\\
              &MM2&0.56(0.11) & 17.14(0.25) & 1.24(0.11) & 20.28(0.25)&-&-& 1.34(0.05) & 18.44(0.02) & 1.56(0.04)\\
              &MM3&-&-&-&-& 1.20(0.09) & 18.24(0.06) & 0.61(0.06) & 17.41(0.05) & 1.50(0.12)\\
              &MM4&-&-&-&-& 0.55(0.07) & 19.42(0.09) & 0.45(0.09) & 19.24(0.04) & 0.80(0.12)\\
G016.93+00.24... &MM1&2.55(0.11) & 23.14(0.25)&-&- &-&-& 1.13(0.05) & 23.77(0.01) & 1.01(0.04)\\
G017.19+00.81... &MM1&1.78(0.14) & 23.68(0.25)& 5.05(0.14) & 25.48(0.25) &-&-& 1.23(0.06) & 25.00(0.02) & 1.50(0.05)\\
              &MM2&4.77(0.20) & 21.93(0.25)& 3.25(0.20) & 24.82(0.25)&-&-& 1.53(0.05) & 22.79(0.02) & 1.78(0.05)\\
              &MM3&4.47(0.12) & 22.18(0.25)& 1.86(0.12) & 24.44(0.25)&-&-& 1.76(0.06) & 22.73(0.02) & 1.54(0.04)\\
              &MM4&0.93(0.05) & 19.89(0.25) & 2.33(0.05) & 22.69(0.25) &-&-& 0.38(0.03) & 22.1(0.05) & 2.19(0.11) \\
G018.26--00.24... &MM1&-&-&-&-& 1.51(0.1) & 65.46(0.08) & 0.78(0.04) & 68.4(0.04) & 2.49(0.09) \\
              &MM2&1.36(0.12) & 65.59(0.25)&-& -&-&-& 1.00(0.05) & 67.79(0.03) & 2.14(0.08)\\
              &MM3&1.16(0.17) & 65.64(0.25)&0.69(0.17) & 71.45(0.25)&-&-& 0.89(0.04) & 68.62(0.04) & 2.67(0.09)\\
              &MM4&0.84(0.17) & 65.26(0.25)&0.94(0.17) & 71.66(0.25)&-&-& 0.72(0.04) & 68.83(0.05) & 2.95(0.11)\\
              &MM5&0.71(0.25) & 67.28(0.25)&-& -&-&-& 0.75(0.04) & 66.56(0.04) & 2.43(0.11)\\
G022.06+00.21... &MM1&0.86(0.20) & 46.54(0.25)& 3.81(0.20) & 52.95(0.25)&-&-& 1.32(0.06) & 51.35(0.03) & 2.01(0.08)\\
              &MM2&0.42(0.10) & 49.35(0.25) & 0.77(0.10) & 53.16(0.25)&-&-& 0.65(0.05) & 51.82(0.05) & 1.75(0.1) \\
G023.28--00.12... &MM1&-&-&-&-& 3.25(0.12) & 99.85(0.03) & 0.77(0.05) & 99.03(0.04) & 2.03(0.09)\\
              &MM2&0.42(0.09) & 96.13(0.25) & 1.63(0.09) & 99.94(0.25) &-&-& 0.32(0.06) & 99.52(0.08) & 1.31(0.18)\\
G024.37--00.15... &MM1&0.71(0.06) & 57.77(0.25) &-&- &-&- & 0.39(0.04) & 59.24(0.08) & 2.49(0.17)\\
              &MM2&-&-&-&-&0.37(0.08) & 53.7(0.24) & 0.45(0.06) & 56.23(0.08) & 2.27(0.22)\\
G024.61--00.33... &MM1&1.08(0.14) & 41.56(0.25) &-&- &-&-& 1.44(0.06) & 42.79(0.02) & 1.72(0.05)\\
              &MM2&1.23(0.09) & 43.52(0.25) &-& -&-&-& 0.70(0.06) & 43.63(0.04) & 1.41(0.09)\\
G024.94--00.15... &MM1&2.62(0.10) & 45.77(0.25) & 1.29(0.10) & 49.28(0.25) &-&-& 0.88(0.05) & 47.37(0.04) & 2.00(0.08)\\
              &MM2&2.06(0.09) & 46.93(0.25) & 0.93(0.09) & 49.91(0.25) &-&-& 0.54(0.04) & 48.17(0.05) & 1.87(0.11)\\
G025.79+00.81... &MM1&0.49(0.08) & 49.18(0.25) & 1.11(0.08) & 51.57(0.25) &-&- & 0.62(0.05) & 49.85(0.04) & 1.60(0.09)\\
              &MM2&-&-&-&-& 0.25(0.04) & 50.27(0.14) & 0.26(0.08) & 50.03(0.11) & 0.97(0.24)\\
G030.90+00.00A..&MM1&-&-&-&-& 2.61(0.14) & 76.66(0.03) & 0.99(0.06) & 75.25(0.04) & 2.14(0.09)\\
G030.90+00.00B.. &MM2&2.43(0.11) & 91.59(0.05) & 1.28(0.11) & 95.06(0.25)&-&-& 0.79(0.05) & 92.81(0.06) & 2.94(0.13)\\
G030.90+00.00C.. &MM3&-&-&-&-& 0.74(0.03) & 93.91(0.12) & 0.52(0.05) & 94.33(0.04) & 1.58(0.12)\\
G034.71--00.63... &MM1&-&-&-&-& 2.38(0.25) & 44.52(0.07) & 1.00(0.05) & 44.76(0.03) & 2.26(0.08)\\ 
              &MM2&2.11(0.11) & 45.23(0.25) &-&-  &-&-& 0.76(0.04) & 45.83(0.03) & 1.73(0.08)\\ 
              &MM3&0.64(0.07) & 48.41(0.25) &-& -&-& -& 0.88(0.05) & 46.49(0.03) & 1.76(0.07)\\
G034.77--00.81... &MM1&-&-&-&-&  0.59(0.07) & 42(0.07) & 0.23(0.03) & 43.68(0.1) & 2.21(0.21) \\
G034.85+00.43... &MM1&-&-&-&-& 0.83(0.07) & 56(0.04) & 0.53(0.06) & 55.72(0.03) & 0.90(0.08)\\ 
G035.49--00.30A.. &MM1&1.14(0.05) & 54.06(0.25) & 0.41(0.05) & 57.7(0.25) &-&-& 0.46(0.04) & 55.61(0.07) & 2.17(0.15)\\
G035.49--00.30B.. &MM2&1.71(0.06) & 44.14(0.25) & 1.06(0.06) & 46.24(0.25)&-&- & 0.65(0.04) & 45.39(0.03) & 1.68(0.07)\\
              &MM3&2.92(0.07) & 45.02(0.25) &-& -&-&-& 0.92(0.05) & 45.85(0.04) & 1.99(0.09)\\ 
G037.44+00.14A.. &MM1&-&-&-&-&- &-& 0.93(0.07) & 18.26(0.02) & 0.64(0.04)\\ 
G037.44+00.14B.. &MM2&-&-&-&-&0.27(0.17) &40.46(0.08)&0.28(0.17)&40.46(0.08)  &1.33(0.17)\\
G050.06+00.06... &MM1&0.98(0.06) & 53.23(0.25) &1.01(0.06) &55.36(0.25) &-&- & 0.62(0.04) & 54.44(0.04) & 1.89(0.09)\\
              &MM2&-&-&-&-& 2.96(0.06) & 54.93(0.01) & 0.60(0.05) & 54.93(0.05) & 1.73(0.11)\\   
G053.81--00.00... &MM1&3.66(0.10) & 23.69(0.25) & 1.21(0.10) & 25.74(0.25) &-&- & 0.76(0.04) & 24.13(0.02) & 1.29(0.06)\\ 
\noalign{\smallskip}
\hline
\end{longtable} 
\tablefoot{Columns are (from left to right) \hcop(1--0) blue peak intensity and its  LSR velocity, \hcop(1--0) red peak intensity and its  LSR velocity, \hcop(1--0) peak intensity and its LRS velocity, \hh13cop(1--0) peak intensity and it LSR velocity, FWHP of the \hh13cop(1--0) line.}  
\end{landscape}  
\twocolumn
}

\onltab{4}{
\onecolumn
\begin{table}
\caption{Properties of \n2hp(1--0) based on Gaussian fits.\label{ta:n2hp10}}
\begin{tabular}{l l c c c c c c c}
\hline\hline
Source name & &$\int T_\mathrm{MB} \Delta v$& $v_\mathrm{LSR}$& $\Delta v$ & $\tau_\mathrm{tot}$ & $T_{\mathrm{ex}}$ & $N_{\mathrm{N_2H^+}}$& $\chi$\\
       & &(K~km~s$^{-1}$) & (km~s$^{-1}$) & (km~s$^{-1}$) &        & (K) & ($10^{12}$~cm$^{-2}$)& ($10^{-9}$)\\
\noalign{\smallskip}
\hline
\noalign{\smallskip}
G012.73--00.58... & MM1&3.7(0.1) & 6.68(0.01) & 0.42(0.01) & 4.11(1.18) & 5.3(1.6) & 5.3(2.2) & 0.26 \\ 
              & MM2&6.7(0.2) & 6.35(0.01) & 0.83(0.04) & 5.97(0.91) & 4.6(0.8) & 12.1(2.8) & -\\ 
G013.28--00.34.. &MM1 & 9.3(0.4) & 41.33(0.02) & 1.68(0.07) & 5.58(0.84) & 4.3(0.8) & 20.6(5.0)& 4.0\\ 
G013.91--00.51... &MM1& 12.9(0.4) & 22.75(0.01) & 1.19(0.03) & 5.34(0.45) & 5.7(0.6) & 22.0(3.0)&1.0\\ 
G013.97--00.45... &MM1 & 8.5(0.3) & 19.87(0.03) & 2.07(0.09) & 1.25(0.47) & 7.8(3.0) & 15.2(8.2)&-\\ 
G014.39--00.75A.. &MM1& 6.6(0.3) & 17.64(0.02) & 1.14(0.07) & 0.78(0.76)  & -&  -&-\\
               &MM2& 6.9(0.3) & 17.12(0.02) & 0.92(0.05) &  0.82(0.86) & -&  -&-\\
G014.39--00.75B.. &MM3& 6.7(0.3) & 21.15(0.01) & 0.72(0.04) & 8.99(1.55) & 4.4(0.9) & 14.8(4.0)&1.0\\ 
G014.63--00.57... &MM1& 45.8(2.3) & 18.45(0.01) & 2.46(0.02) & 2.35(0.13) & 13.3(0.8) & 89.0(7.4)&0.52\\ 
               &MM2&  18.0(0.7) & 18.23(0.01) & 1.27(0.03) & 3.90(0.42) & 7.6(0.9) & 27.8(4.5)&0.5\\ 
               &MM3& 7.7(0.3) & 17.44(0.03) & 1.34(0.08) & 3.71(0.93) & 4.8(1.3) & 13.0(4.8)&1.4\\ 
               &MM4& 2.8(0.2) & 19.03(0.03) & 0.63(0.06) & 6.66(2.54) & 3.7(1.6) & 7.3(4.2)&2.4\\ 
G016.93+00.24...&MM1 & 8.7(0.3) & 23.63(0.02) & 0.84(0.05) & 4.17(1.05) & 5.9(1.6) & 12.8(4.8)&1.0\\ 
G017.19+00.81...&MM1 & 21.0(0.6) & 24.83(0.01) & 1.54(0.04) & 0.43(0.37) &-&-&-\\
               &MM2 & 32.9(1.1) & 22.65(0.01) & 2.21(0.04) & 1.0(0.23) & 27.7(6.4) & 135.0(43.8)&-\\
              &MM3 & 25.1(0.7) & 22.50(0.01) & 1.74(0.02) & 0.94(0.17) & 29.(0.3) & 109.7(28.0)&-\\
              &MM4 & 4.5(0.2) & 22.00(0.03) & 2.10(0.05) & 0.44(0.45) &-&-&- \\
G018.26--00.24...&MM1 & 23.5(1.0) & 68.45(0.02) & 2.50(0.06) & 2.10(0.32) & 8.7(1.4) & 37.4(8.3)&1.1\\ 
               &MM2& 26.0(1.0) & 67.87(0.02) & 2.16(0.01) & 4.00(0.22) & 7.1(0.4) & 43.2(3.4)&1.7\\ 
               &MM3 & 37.4(1.3) & 68.65(0.01) & 2.70(0.01) & 0.18(0.01)&-&  -&-\\
               &MM4 & 30.6(1.0) & 68.72(0.01) & 2.94(0.02) &  0.18(0.21)& -&  -&-\\
               &MM5 & 16.8(0.8) & 66.24(0.02) & 1.75(0.06) & 3.47(0.45) & 6.4(0.9) & 25.5(5.0)&1.6\\ 
G022.06+00.21...&MM1 & 28.7(1.3) & 51.25(0.01) & 2.50(0.01) & 1.17(0.01) & 17.7(0.1) & 76.6(0.5)&0.48\\ 
                &MM2 & 8.3(0.4) & 51.59(0.02) & 1.74(0.08) & 1.57(0.52) & 7(2.4) & 13.4(6.4)&0.52\\ 
G023.28--00.12...&MM1 & 17.9(0.6) & 98.87(0.02) & 2.11(0.06) & 0.45(0.32) & -&  -&-\\
               &MM2 & 4.07(0.2) & 99.35(0.03) & 0.95(0.09) & 4.95(1.98) & 3.9(1.8) & 8.9(5.5)&-\\ 
G024.37--00.15...&MM1 & 13.5(0.8) & 58.87(0.02) & 2.53(0.05) & 1.33(0.27) & 8.6(1.8) & 23.5(6.9)&1.4\\ 
               &MM2& 12.8(0.5) & 56.18(0.02) & 2.28(0.07) & 1.55(0.32) & 7.9(1.7) & 21.4(6.4)&1.3\\ 
G024.61--00.33...&MM1 & 16.2(0.5) & 42.82(0.01) & 2.12(0.04) & 0.19(0.54)& -&  -&-\\
                &MM2& 10.6(0.4) & 43.56(0.01) & 1.66(0.04) & 0.18(0.06) &  -&  -&-\\
G024.94--00.15...&MM1 & 20.0(0.8) & 47.24(0.02) & 2.48(0.05) & 1.30(0.20) & 12.1(1.8) & 41.6(8.9)&0.88\\ 
                &MM2 & 11.9(0.4) & 48.22(0.01) & 1.92(0.05) & 0.36(0.30) & - & -&-\\ 
G025.79+00.81... &MM1& 11.1(0.3) & 49.78(0.01) & 1.56(0.05) & 1.35(0.34) & 10.4(2.9) & 20.7(8.2)&-\\ 
                 &MM2& 3.0(0.2) & 49.78(0.06) & 1.90(0.24) &1.62(0.91) & -&  -&-\\
G030.90+00.00A.. &MM1& 22.8(0.8) & 75.21(0.01) & 2.38(0.02) & 0.18(0.10) & -&  -&-\\
G030.90+00.00B.. &MM2& 25.2(0.3) & 92.95(0.01) & 3.15(0.02) &  0.26(0.12) & -&  -&-\\
G030.90+00.00C.. &MM3& 13.6(0.6) & 94.29(0.02) & 1.50(0.06) & 1.76(0.51) & 9.6(2.9) & 22.4(9.4)&1.1\\ 
G034.71--00.63... &MM1& 11.7(0.7) & 44.57(0.04) & 2.31(0.10) & 2.61(0.61) & 5.3(1.4) & 18.5(6.6)&1.7\\ 
                 &MM2& 11.5(0.5) & 45.54(0.02) & 1.58(0.08) & 4.90(0.91) & 4.9(1.1) & 20.9(6.2)&1.5\\ 
                                &MM3& 8.9(0.3) & 46.24(0.01) & 1.59(0.05) & 1.47(0.35) & 8.1(2) & 14.7(5.1)&1.7\\ 
G034.77--00.81...&MM1 & 1.6(0.1) & 43.29(0.06) & 1.60(0.19) &1.11(1.26) &  -&  -&-\\
G034.85+00.43...&MM1 & 3.5(0.1) & 55.68(0.02) & 0.72(0.05) & 8.20(2.04) & 3.6(1.1) & 9.8(3.9)&-\\ 
G035.49--00.30A..&MM1 & 8.3(0.3) & 55.28(0.01) & 2.14(0.04) & 0.18(0.02) & - & -&-\\ 
G035.49--00.30B..&MM2 & 9.4(0.4) & 45.40(0.01) & 1.37(0.04) & 5.52(0.64) & 4.6(0.6) & 18.4(3.3)&1.3\\ 
                &MM3& 14.8(0.5) & 45.69(0.01) & 1.86(0.04) & 3.28(0.30) & 6(0.6) & 23.0(3.2)&1.7\\ 
G037.44+00.14A.. &MM1& 3.4(0.1) & 18.14(0.01) & 0.57(0.02) & 4.19(0.90) & 4.5(1) & 5.6(1.7)&-\\ 
G037.44+00.14B.. &MM2& 1.0(0.1) & 40.15(0.13) & 1.07(0.35) & 0.92(5.20)& -&  -&-\\
G050.06+00.06... &MM1 & 5.3(0.2) & 54.23(0.02) & 1.32(0.07) & 3.83(0.82) & 4.1(1) & 10.3(3.4)&1.4\\ 
                 &MM2 & 4.3(0.2) & 54.93(0.03) & 1.32(0.08) & 1.85(0.85) & 5(2.4) & 6.8(4.5)&0.63\\ 
G053.81--00.00... &MM1 & 9.0(0.3) & 24.11(0.01) & 1.64(0.03) & 0.45(0.25) & - & - &- \\ 
\noalign{\smallskip}
\hline
\end{tabular}
\tablefoot{Columns are (from left to right) source name, n2hp(1--0) peak main beam brightness temperature, LSR velocity at the \n2hp(1--0) peak intensity, FWHP  \hcop(1--0) line width, \n2hp(1--0) total optical depth, \n2hp(1--0) excitation temperature, \n2hp\ column density, and the \n2hp\ abundance relative to the H$_2$ column density obtained from the 1.2\,mm observations (Paper I).}
\end{table}
\twocolumn
}

\onltab{5}{
\onecolumn
\begin{landscape}
\setlength{\LTcapwidth}{1.4\textheight}
\begin{longtable}{llcccccccccc}
\caption{Observed properties of \n2hp(3--2) and \hcop(4--3). When the source was not observed the field is marked as `..'. \label{ta:apex1}}\\
\hline\hline
Source name & &\multicolumn{3}{c}{\underline{\hspace{2cm}\n2hp(3--2)\hspace{2cm}}} & \multicolumn{7}{c}{\underline{\hspace{5.5cm}\hcop(4--3)\hspace{5.5cm}}}\\
       & &$\int T_{\mathrm{MB}}\mathrm{d}v$& $v_\mathrm{LSR}$ & $\Delta v$ &$\int T_{\mathrm{MB}}\mathrm{d}v$& $v_\mathrm{LSR}$ & $\Delta v$&$T_\mathrm{blue}$ &$ v_\mathrm{LSR, blue}$  & $T_\mathrm{red}$ &$ v_\mathrm{LSR, red}$ \\
       & & (K~km~s$^{-1}$) & (km~s$^{-1}$) & (km~s$^{-1}$)  & (K~km~s$^{-1}$) & (km~s$^{-1}$) & (km~s$^{-1}$) & (K) & (km~s$^{-1}$) & (km~s$^{-1}$)\\
\noalign{\smallskip}
\hline
\noalign{\smallskip}
G013.91--00.51.. &MM1   & 2.4(0.2) & 22.77(0.08) &1.56(0.2) &4.2(0.4) & 22.43(0.06) & 1.38(0.16) & - & - & - & -\\
G014.39--00.75A. &MM1& 1.9(0.3) & 17.67(0.1) & 1.32(0.26)& ..&.. &..& .. & .. & .. & ..\\ 
G014.63--00.57.. &MM1& 25.6(0.4) & 18.63(0.04) & 5.34(0.08) & 27.9(0.4) & 17.75(0.04) & 5.05(0.1)  & 6.92(0.30) & 17.1(0.2)& 4.68(0.30) & 20.4(0.2) \\ 
              &MM2& 8.2(0.4)  & 18.42(0.07) & 3.31(0.2)  & 7.6(0.3)  & 18.29(0.08) & 3.75(0.19) & 1.99(0.26)& 17.3(0.2) & 1.78(0.26) & 18.9(0.2)\\ 
G017.19+00.81.. &MM1& 9.8(0.4)  & 25.18(0.05) & 2.88(0.14) & 8.8(0.5)   & 25.05(0.06) & 2.31(0.14) & -        & -        & 3.50(0.45) & 25.5(0.2)\\ 
               &MM2& 18.7(0.4) & 22.82(0.04) & 4.21(0.1)  & 25.5(0.7) & 22.78(0.08) & 5.99(0.19) & 3.92(0.45)& 21.4(0.2) & 5.26(0.45) & 23.8(0.2)\\
               &MM4& 1.7(0.3)  & 21.57(0.15) & 1.88(0.37) & 4.1(0.5)  & 21.97(0.15) & 2.61(0.36) & -& - & 1.4(0.44) & 22.7(0.2)\\ 
G018.26--00.24.. &MM1& 8.6(0.3)  & 68.52(0.10) & 5.05(0.21) & 9.2(0.7)  & 68.67(0.21) & 5.38(0.48) & 1.55(0.46)& 66.9(0.2) & 2.12(0.46) & 69.4(0.2)\\ 
               &MM2& 7.6(0.4)  & 67.95(0.10) & 4.37(0.25) & 3.6(0.3)  & 67.76(0.15) & 3.15(0.28) & -& - & - & -\\ 
               &MM3& 12.9(0.3) & 68.73(0.06) & 4.33(0.14) & 11.7(0.5) & 68.52(0.09) & 4.21(0.21) & -& - & -          & -        \\  
              &MM4& 8.9(0.4)  & 68.90(0.09) & 4.42(0.20) & 7.2(0.5)  & 68.84(0.13) & 4.26(0.32) & -         & -         & - & -\\
G022.06+00.21.. &MM1& 16.3(0.4) & 51.24(0.06) & 5.33(0.15) &27.0(0.9)  & 51.12(0.09) & 5.31(0.22) & - & - & - &-\\ 
G023.28--00.12.. &MM1& 4.7(0.4)  & 98.96(0.14) & 4.01(0.40) & 5.1(0.6)  & 98.57(0.18) & 3.03(0.46) & 2.01(0.51)& 97.8(0.2) & -&-\\ 
G024.37--00.15.. &MM2 & 5.4(0.3) & 56.21(0.11) & 3.92(0.27) & 6.1(0.7)  & 57.11(0.39) & 5.48(0.86) & -         & -         &    -       & - \\ 
G024.61--00.33.. &MM1 & 6.7(0.3) & 42.86(0.07) & 3.49(0.19) & 9.8(0.6)  & 42.80(0.10) & 3.10(0.24) & -         & -         & 3.08(0.54) & 43.5(0.2)\\ 
G024.94--00.15.. &MM1 & 7.4(0.3) & 47.13(0.09) & 3.94(0.22) & 8.7(0.6)  & 46.70(0.17) & 4.67(0.35) & 2.20(0.44)& 45.9(0.2) & 2.01(0.44) & 47.9(0.2)\\ 
G030.90+00.00A. &MM1 & 7.7(0.3)  & 75.30(0.07) & 3.73(0.17) & 7.7(0.7) & 75.17(0.13)  & 3.14(0.34) & -         &  -        &   -        &  -        \\
G030.90+00.00B. &MM2& 6.7(0.3)  & 92.96(0.11) & 4.69(0.26) & 5.2(0.6) & 92.53(0.19)  & 3.54(0.54) & -         &  -        &   -        &  -         \\
G034.71--00.63.. &MM1& 6.1(0.4)  & 44.76(0.12) & 4.40(0.29) & 16.6(0.9) & 44.38(0.15) & 6.16(0.41) & 3.15(0.50)& 43.8(0.2) & 3.17(0.50) & 45.8(0.2)\\
                &MM2& 2.7(0.4)  & 45.88(0.29) & 4.09(0.62) & 1.2(0.5)  & 45.14(0.13) & 0.74(0.44) & -         &  -        &   -        &  -         \\
G053.81--00.00 &MM1& 3.7(0.34) & 24.44(0.11) & 2.48(0.32) & .. & .. & ..& .. & .. & .. & ..\\
\noalign{\smallskip}
\hline
\end{longtable}
\tablefoot{Columns are (from left to right) source name, \n2hp(3--2) peak main beam brightness temperature, LSR velocity at the \n2hp(3--2) peak intensity, FWHP  \hcop(4--3) line width, \n2hp(3--2) peak main beam brightness temperature, LSR velocity at the \hcop(4--3) peak intensity, FWHP  \hcop(4--3) line width, blue \hcop(4--3) line intensity peak and its LRS velocity, and red \hcop(4--3) line intensity peak and its LSR velocity.}
\end{landscape}
\twocolumn
}

\onltab{6}{
\begin{table*}
\centering
\setlength{\LTcapwidth}{1.4\textheight}
\caption{Star formation tracers (ii): $K$ levels of \mecn(5--4). $K$-level detections are marked as `$\surd$', non-detections as `-'. \label{ta:mecn}}
\begin{tabular}{llcccc c c c}
\hline\hline
\noalign{\smallskip}
Source name        &  &  \multicolumn{4}{c}{$K$ levels}& $T_\mathrm{rot}$(\mecn) & $T_\mathrm{rot}$(\amm) & N(\mecn)\\
              &  &  0 & 1 & 2 & 3          &  (K)              &    (K)  &  ($10^{12}$cm$^{-2}$) \\
\noalign{\smallskip}
\hline
\noalign{\smallskip}
G014.63--00.57... &MM1& $\surd$ &$\surd$ & $\surd$ & $\surd$& 29(6) & 18.1& 5.9(2.6)\\
              &MM2& $\surd$ &$\surd$ & - &- & - & 15.7& - \\
G017.19+00.81... &MM2& $\surd$ &$\surd$ & $\surd$ &  -    &41(23) & 18.7&\\  
              &MM3& $\surd$ &$\surd$ & -       & -       &  -& 20.0&-\\
              &MM4& $\surd$ &$\surd$ & - & -       &- &  20.1& -\\
G018.26--00.24... &MM1& $\surd$ &$\surd$ & -       & -       &-& 18.2&- \\
              &MM3& $\surd$ &$\surd$ & $\surd$ & -       &23(10) & 15.7&-\\
              &MM4& $\surd$ &$\surd$ & -       & - &- & 16.6&-\\
G022.06+00.21... &MM1& $\surd$ &$\surd$ & $\surd$ & $\surd$ &32(5) & 24.7&9.7(2.8)\\
G023.28--00.12... &MM1& $\surd$ &$\surd$ & -       & -       & -& 18.1&- \\ 
G024.37--00.15... &MM2& $\surd$ &$\surd$ &- & -       & -& 15.7&-\\ 
G024.61--00.33... &MM1& $\surd$ &$\surd$ & $\surd$   & $\surd$ &68(28) &17.5& 5.9(4.7)\\
G024.94--00.15... &MM1& $\surd$ &$\surd$ & $\surd$ & -       &39(31) & 15.2&- \\
              &MM2& $\surd$ &$\surd$ & - & -       &   -   & 15.2&-\\
G034.71--00.63... &MM1& $\surd$ &$\surd$ & $\surd$ & $\surd$ &29(6) &17.8& 5.4(2.8)\\
G035.49--00.30A.. &MM1& $\surd$ & $\surd$     & -       & -       &-  & 13.2& -\\
\noalign{\smallskip}
\hline
\end{tabular}
\tablefoot{Columns are (from left to right) source name, detection of \mecn\ $K$ levels, \mecn\ rotational temperature, \amm\ rotational temperature, and \mecn\ column density.}
\end{table*}
}

\onltab{7}{
\begin{table*}
\centering
\caption{Star formation tracers (i): SiO(2--1) and H$_2$CO(4$_{03}$--3$_{04}$) based on Gaussian fits. Non detections are marked by `-', see Table~\ref{ta:det} for upper limits.\label{ta:sf}}
\begin{tabular}{llcccccc}
\hline\hline
Source name        &  & \multicolumn{3}{c}{SiO(2--1)} & \multicolumn{3}{c}{H$_2$CO(4--3)}\\
              &  & $T_{MB}$ & $v_\mathrm{LSR}$  & $\Delta V$ & $T_{MB}$ & $v_\mathrm{LSR}$  & $\Delta V$\\
              &  & (K) &  (km~s$^{-1}$) & (km~s$^{-1}$) &(K) &  (km~s$^{-1}$) & (km~s$^{-1}$)\\
\noalign{\smallskip}
\hline
\noalign{\smallskip}
G013.28--00.34... &MM1&  0.11(0.03) & 39.47(0.43) & 4.74(0.88)&- &- &- \\
G013.91--00.51... &MM1&   -          &  -           &   -      & 0.53(0.17) & 22.94(0.13) &1.24(0.3)\\                                          
G014.39--00.75A.. &MM1&   -          & -            &    -       & 0.80(0.30) & 17.64(0.14)& 1.01(0.28) \\                                     
G014.63--00.57... &MM1&  0.45(0.04) & 17.60(0.10)   & 5.17(0.34)& 2.41(0.10) & 17.23(0.05) &3.62(0.11)\\ 
              &MM2&  0.08(0.02) & 18.73(0.53) & 6.42(1.28)& 0.85(0.15) & 18.24(0.12)& 2.71(0.37)\\
G017.19+00.81... &MM1&  0.09(0.02) & 24.68(0.50)  & 7.77(1.05 &0.56(0.13) & 24.96(0.17)& 2.28(0.39)\\  
              &MM2&  0.79(0.11)$^a$ & 22.22(0.25)\footnote{Not based on Gaussian fit.} & 26.48(0.64)&1.39(0.10) & 22.46(0.12) &5.92(0.34)\\ 
              &MM3&  0.76(0.11)$^a$ & 21.95(0.25)$^a$ & 19.42(0.60)&- & -&-\\ 
G018.26--00.24... &MM1&  0.36(0.02) & 66.88(0.16)& 9.18(0.47)& 0.69(0.12) & 68.02(0.18)& 3.69(0.49)\\
              &MM2&  0.09(0.02) & 68.01(0.64)& 11.14(2.35)& 0.41(0.12) & 67.74(0.29)& 3.12(0.70)\\
              &MM3&  0.21(0.02) & 68.50(0.27) &10.06(0.85)& 0.46(0.14) & 68.52(0.22)& 3.06(0.75)\\
              &MM4&  0.25(0.02) & 69.24(0.28)& 13.02(0.83)& 0.34(0.14) & 69.91(0.43)& 4.30(1.46)\\
G022.06+00.21... &MM1&  0.24(0.02) & 48.27(0.35) &16.26(0.89)& 1.88(0.11) & 51.04(0.08)& 4.78(0.22)\\
G023.28--00.12... &MM1&  0.13(0.02) &101.05(0.55) &12.37(1.39)&- &- &- \\
G024.37--00.15... &MM2&  0.25(0.02) & 54.53(0.30) &10.81(0.84)& 0.69(0.11) & 55.79(0.21)& 5.13(0.64)\\
G024.61--00.33... &MM1&  0.11(0.02) & 40.44(0.49) &10.18(1.2) &0.54(0.10) & 42.96(0.24) &3.63(0.50)\\
G024.94--00.15... &MM1&  0.24(0.03) & 47.14(0.19) &5.91(0.51) & 0.69(0.12) & 46.83(0.19)& 3.87(0.50)\\
              &MM2&  0.18(0.02) & 47.39(0.38) &11.61(1.08)& -& -&-\\
G025.79+00.81... &MM1&  0.12(0.04) & 49.20(0.56) &8.04(2.01) &- &- &-\\
G034.71--00.63... &MM1&  0.15(0.02) & 44.40(0.46) &14.14(1.19)& 1.49(0.13) & 44.12(0.08) &3.02(0.21)\\
              &MM2&  0.35(0.08) & 45.47(0.23) & 6.63(1.25)& 0.27(0.09) & 45.26(0.57) &5.03(1.31)\\
              &MM3&  0.17(0.04) & 45.94(0.20) &3.37(0.62) &- &- &-\\
G035.49--00.30B.. &MM2&  0.14(0.03) & 45.07(0.25) &3.88(0.67)&- &- &-\\ 
              &MM3&  0.18(0.05) & 45.32(0.14) &1.57(0.32)&- &- &-\\ 
G053.81--00.00... &MM1&  0.11(0.01) & 23.14(0.24) &9.14(0.52)&0.47(0.11) & 23.79(0.39) &4.81(0.79)\\
\noalign{\smallskip}
\hline
\end{tabular}
\tablefoot{Columns are (from left to right) source name, SiO peak main beam brightness temperature, LSR velocity at the SiO peak intensity, FWHP SiO line width, \h2co\ peak main beam brightness temperature, LSR velocity at the \h2co\ peak intensity, and FWHP \h2co\ line width. }
\end{table*}
}

\onltab{8}{
\onecolumn
\begin{table}
\centering
\caption{Observed properties of \c18o(2--1) based on Gaussian fits. Non detections are marked by `-', see Table~\ref{ta:det} for upper limits.\label{ta:c18o-ii}}
\begin{tabular}{llccc}
\hline\hline
\noalign{\smallskip}
Source name & & $\int T_{\mathrm{MB}}\mathrm{d}v$& $v_\mathrm{LSR}$ & $\Delta v$ \\
       & & (K~km~s$^{-1}$) & (km~s$^{-1}$) & (km~s$^{-1}$)  \\
\noalign{\smallskip}
\hline
\noalign{\smallskip}
G012.73--00.58... &MM1 & - & - & - \\
              &MM2 & 2.1(0.7) & 6.14(0.44) & 1.84(0.7) \\
G013.28--00.34... &MM1 & 4.3(0.5) & 41.09(0.12) & 1.53(1.79) \\
G013.91--00.51... &MM1 & 9.7(0.5) & 22.95(0.07) & 2.81(0.19) \\
G013.97--00.45... &MM1 & 12.4(0.5) & 19.66(0.05) & 2.57(0.12) \\
G014.39--00.75A.. &MM1 & 12.9(0.5) & 17.67(0.04) & 2.10(0.10) \\
              &MM2 & 14.1(0.5) & 17.6(0.04) & 2.15(0.08) \\
G014.39--00.75B..  &MM3 & 5.8(0.3) & 21.06(0.04) & 1.81(0.28) \\
G014.63--00.57... &MM1 & 18.8(0.2) & 18.02(0.02) & 3.14(0.05) \\
              &MM2 & 17.1(0.3) & 18.34(0.02) & 2.77(0.05) \\
              &MM3 & 11.0(0.2) & 17.36(0.02) & 2.24(0.04) \\
              &MM4 & 6.2(0.2) & 18.9(0.03) & 1.96(0.14) \\
G016.93+00.24... &MM1 & 8.7(0.3) & 23.55(0.04) & 1.99(0.08) \\
G017.19+00.81... &MM1 & 9.9(0.2) & 24.23(0.02) & 2.85(0.06) \\
              &MM2 & 9.6(0.2) & 22.63(0.02) & 2.41(0.05) \\
              &MM3 & 9.0(0.1) & 22.75(0.02) & 2.51(0.05) \\
              &MM4 & 12.5(0.1) & 21.72(0.01) & 2.66(0.04) \\
G018.26--00.24... &MM1 & 10.3(0.5) & 67.82(0.09) & 3.70(0.25) \\
              &MM2 & 18.2(0.4) & 67.44(0.03) & 2.86(0.08) \\
              &MM3 & 15.5(0.4) & 68.51(0.05) & 3.72(0.11) \\
              &MM4 & 13.3(0.7) & 68.4(0.1) & 3.78(0.22) \\
              &MM5 & 17.0(0.5) & 66.48(0.04) & 2.82(0.08) \\
G022.06+00.21... &MM1 & 19.0(0.3) & 51.29(0.02) & 3.08(0.05) \\
              &MM2 & 9.6(0.4) & 51.87(0.05) & 2.69(0.12) \\
G023.28--00.12... &MM1 & 14.7(0.9) & 99.15(0.08) & 2.89(0.22) \\
              &MM2 & 6.1(0.8) & 99.54(0.15) & 1.96(0.27) \\
G024.37--00.15... &MM1 & 14.6(0.4) & 58.99(0.04) & 2.93(0.1) \\
              &MM2 & 4.6(0.5) & 56.32(0.2) & 3.67(0.44) \\
G024.61--00.33... &MM1 & 12.9(0.2) & 42.43(0.02) & 2.45(0.04) \\
              &MM2 & 3.9(0.3) & 43.43(0.09) & 3.12(0.26) \\
G024.94--00.15... &MM1 & 5.5(0.3) & 47.07(0.05) & 2.48(0.16) \\
              &MM2 & 4.7(0.3) & 47.74(0.1) & 3.26(0.2) \\
G025.79+00.81... &MM1 & 3.4(0.2) & 49.8(0.06) & 2.16(0.19) \\
              &MM2 & 4.2(0.2) & 49.04(0.05) & 2.77(0.12) \\
G030.90+00.00A..&MM1 & 19.8(0.7) & 75.11(0.05) & 3.22(0.13) \\
G030.90+00.00B..&MM2 & 17.8(0.4) & 92.64(0.04) & 3.86(0.10) \\
G030.90+00.00C..&MM3 & 15.4(0.5) & 94.35(0.08) & 4.46(0.18) \\
G034.71--00.63... &MM1 & 7.9(0.2) & 44.93(0.04) & 3.31(0.12) \\
              &MM2 & 5.7(0.2) & 45.96(0.06) & 2.82(0.14) \\
              &MM3 & 6.3(0.2) & 46.55(0.04) & 2.66(0.09) \\
G034.77--00.81... &MM1 & 5.2(0.3) & 43.72(0.08) & 3.44(0.21) \\
G034.85+00.43... &MM1 & 2.8(0.2) & 55.51(0.07) & 2.11(0.14) \\
G035.49--00.30A.. &MM1 & 11.8(0.2) & 55.4(0.03) & 2.77(0.06) \\
G035.49--00.30B.. &MM2 & 4.1(0.2) & 45.16(0.04) & 2.04(0.13) \\
              &MM3 & 5.7(0.2) & 45.67(0.05) & 2.46(0.11) \\
G037.44+00.14A.. &MM1 & 1.0(0.1) & 18.07(0.05) & 1.37(1.5) \\
G037.44+00.14B.. &MM2 & 8.0(0.2) & 40.11(0.03) & 1.87(0.05) \\
G050.06+00.06... &MM1 & 10.1(0.2) & 54.19(0.02) & 2.27(0.05) \\
              &MM2 & 10.4(0.3) & 54.82(0.03) & 2.06(0.05) \\
G053.81--00.00... &MM1 & 3.8(0.1) & 23.89(0.02) & 2.30(0.07) \\
\noalign{\smallskip}
\hline
\end{tabular}
\tablefoot{Column are (from left to right) source name, integrated line intensity, LSR velocity at the line intensity peak, and FWHP line width. }
\end{table}
\twocolumn
}

\onltab{9}{
\onecolumn
\begin{table}
\centering
\caption{Derived abundance and depletion from the \c18o(2--1) line. \label{ta:c18o}}
\begin{tabular}{llccccc}
\hline\hline
\noalign{\smallskip}
Source name & & $T_\mathrm{rot}$(\amm ) & $N_\mathrm{C^{18}O}$ & $N_\mathrm{H_2}$ & $\chi_\mathrm{C^{18}O}$& $\eta$\\
       & & (K)            & ($10^{15}\,\mathrm{cm}^{-2}$) & ($10^{22}\,\mathrm{cm}^{-2}$) & ($10^{-8}$)&\\
\noalign{\smallskip}
\hline
\noalign{\smallskip}
G012.73--00.58... &MM1& 9.3 &-  & 4.5 & - &-  \\ 
              &MM2& 11.4 & 1.3(0.5) & - &  -&  -\\ 
G013.28--00.34... &MM1&14.9 & 2.4(0.3) & 2.9 & 8.3 & 2.1 \\ 
G013.91--00.51... &MM1&14.1 & 5.5(0.3) & 7.7 & 7.1 & 2.4 \\ 
G013.97--00.45... &MM1&16.6 & 6.9(0.3) & - &-  &-  \\ 
G014.39--00.75A.. &MM1& 17.4 & 7.2(0.3) & 5.7 & 12.6 & 1.3 \\ 
              &MM2& 20.0 & 7.9(0.3) &  -&  -& - \\ 
G014.39--00.75B..               &MM3& 11.8 & 3.5(0.2) & 5.7 & 6.2 & 2.8 \\ 
G014.63--00.57... &MM1& 18.1 & 10.5(0.1) & 26.3 & 4 & 4.3 \\ 
              &MM2& 15.7 & 9.6(0.2) & 15.4 & 6.2 & 2.7 \\ 
              &MM3& 15.8 & 6.1(0.1) & 5.2 & 11.8 & 1.4 \\ 
              &MM4& 19.1 & 3.4(0.1) & 3 & 11.5 & 1.5 \\ 
G016.93+00.24... &MM1& 14.0 & 4.9(0.2) & 4.2 & 11.8 & 1.4 \\ 
G017.19+00.81... &MM1& 17.2 & 5.5(0.1) & 4.9 & 11.3 & 1.5 \\ 
              &MM2& 18.7 & 5.4(0.1) & 17.4 & 3.1 & 5.5 \\ 
              &MM3& 20.0 & 5.1(0.1) & 5.3 & 9.6 & 1.8 \\ 
              &MM4& 20.1 & 7.0(0.1) & 2.9 & 24.3 & 0.7 \\ 
G018.26--00.24... &MM1& 18.2 & 5.7(0.3) & 9.8 & 5.9 & 2.9 \\ 
              &MM2& 17.4 & 10.1(0.2) & 8.9 & 11.4 & 1.5 \\ 
              &MM3& 15.7 & 8.7(0.2) & 7.1 & 12.2 & 1.4 \\ 
              &MM4& 16.6 & 7.4(0.4) & 5.7 & 12.9 & 1.3 \\ 
              &MM5& 16.8 & 9.5(0.2) & 6 & 15.8 & 1.1 \\ 
G022.06+00.21... &MM1& 24.7 & 11.2(0.2) & 25.4 & 4.4 & 3.8 \\ 
              &MM2& 15.5 & 5.4(0.2) & 7.9 & 6.8 & 2.5 \\ 
G023.28--00.12... &MM1& 18.1 & 8.2(0.5) & 6.5 & 12.6 & 1.3 \\ 
              &MM2& 10.0 & 3.4(0.6) &-  &-  &-  \\ 
G024.37--00.15... &MM1& 18.6 & 8.1(0.2) & 4.9 & 16.6 & 1 \\ 
              &MM2& 15.7 & 2.6(0.3) & 4.4 & 5.8 & 2.9 \\ 
G024.61--00.33... &MM1& 17.5 & 7.2(0.1) & 5.9 & 12.2 & 1.4 \\ 
              &MM2& 15.5 & 2.2(0.1) & 3.5 & 6.3 & 2.7 \\ 
G024.94--00.15... &MM1& 15.2 & 3.1(0.1) & 6.5 & 4.7 & 3.6 \\ 
              &MM2& 15.2 & 2.7(0.1) & 5.5 & 4.8 & 3.5 \\ 
G025.79+00.81... &MM1& 13.1 & 1.9(0.1) &-  &-  &-  \\ 
              &MM2& 15.0 & 2.4(0.1) & - & - & -\\ 
G030.90+00.00A..&MM1& 18.6 & 11(0.4) & 5.2 & 21.2 & 0.8 \\ 
G030.90+00.00B..              &MM2& 15.0 & 10(0.2) & - & - & -\\ 
G030.90+00.00C..              &MM3& 16.4 & 8.6(0.3) & 4.3 & 20 & 0.9 \\ 
G034.71--00.63... &MM1& 17.8 & 4.4(0.1) & 5.8 & 7.6 & 2.2 \\ 
              &MM2& 12.4 & 3.4(0.1) & 5.2 & 6.5 & 2.6 \\ 
              &MM3& 17.1 & 3.5(0.1) & 2.5 & 14 & 1.2 \\ 
G034.77--00.81... &MM1& 15.0 & 2.9(0.1) & - & - &-  \\ 
G034.85+00.43... &MM1& 13.2 & 1.6(0.1) &-  & - & - \\ 
G035.49--00.30A.. &MM1& 18.6 & 6.6(0.1) & 6.6 & 10 & 1.7 \\ 
G035.49--00.30B..              &MM2& 11.9 & 2.5(0.1) & 5.3 & 4.6 & 3.7 \\ 
              &MM3& 13.6 & 3.2(0.1) & 4.2 & 7.7 & 2.2 \\ 
G037.44+00.14A.. &MM1& 15.0 & 0.6(0.05) & - &-  & - \\ 
G037.44+00.14B..              &MM2& 15.0 & 4.5(0.1) &  -&-  & -  \\ 
G050.06+00.06... &MM1& 14.9 & 5.6(0.1) & 4.4 & 12.8 & 1.3 \\ 
              &MM2& 14.1 & 5.9(0.1) & 4.6 & 12.9 & 1.3 \\ 
G053.81--00.00... &MM1& 12.4 & 2.2(0.1) & 6.9 & 3.2 & 5.3 \\ 
\noalign{\smallskip}
\hline
\end{tabular}
\tablefoot{Columns are (from left to right) source name, \amm\ rotational temperature, \c18o\ column density, hydrogen column density from the 1.2\,mm emission (Paper I), \c18o\ abundance, and CO depletion factor.}
\end{table}
\twocolumn
}

\end{document}